\documentclass[preprint2]{aastex}

\slugcomment{Submitted to AJ}

\shorttitle{Small Scale Systems of Galaxies. III}
\shortauthors{Gr\" utzbauch et al.}

\begin{document}

\title{
Small Scale Systems of Galaxies. III. \\
X-ray detected E+S galaxy pairs in low density environments}

\author{R.~Gr\"utzbauch$^{1}$, G.~Trinchieri$^{2}$, R.~Rampazzo$^{3}$,
E.V.~Held$^{3}$\\
L.~Rizzi$^{4}$, J.W.~Sulentic$^{5}$ and W.W.~Zeilinger$^{1}$}
\affil{$^{1}$ Institut f\"ur Astronomie, Universit\"at Wien,
T\"urkenschanzstra{\ss}e 17, A-1180 Wien, Austria \\
         $^{2}$ INAF - Osservatorio Astronomico di Brera, Via Brera 28,
I-20121, Milano, Italy\\
        $^{3}$INAF - Osservatorio Astronomico di Padova, Vicolo
        dell'Osservatorio 5,I-35122, Padova, Italy\\
        $^{4}$Institut for Astronomy, University of Hawaii, Woodlown Drive,
        Honolulu, HI 96822, USA\\
        $^{5}$ Department of Physics and Astronomy, University of Alabama,
 Tuscaloosa, AL 35487, USA
}
\email{$^{1}$gruetzbauch@astro.univie.ac.at, $^{2}$
ginevra@brera.mi.astro.it,\\
$^{3}$  roberto.rampazzo@oapd.inaf.it, $^{3}$ enrico.held@oapd.inaf.it, $^{4}$
rizzi@ifa.hawaii.edu\\
$^{5}$ giacomo@merlot.astr.ua.edu, $^{1}$ zeilinger@astro.univie.ac.at}

\begin{abstract}
We present a comprehensive study of the local environments of four E+S
galaxy pairs with the main goal to investigate their
formation/evolution histories.

We obtained new XMM-Newton data for two pairs (RR~143 and RR~242) that
complements existing ROSAT data for the other two (RR~210 and RR~216).
The new observations reveal diffuse X-ray emission in both pairs with
luminosities of L$_X \sim 3 \times 10^{41}$ erg~s$^{-1}$ (0.5-2.0
keV).  The emission is asymmetric in both cases and extends out to
$\sim$ 500\arcsec\ (120 kpc) and 700\arcsec\ (160 kpc) in RR~143 and
RR~242 respectively.  The nucleus of RR~242 hosts a low luminosity
(L$_X \sim 10^{41}$ erg s$^{-1}$, 2.0-10 keV) mildly absorbed (N$_H
\sim 10^{22}$ cm$^{-2}$) AGN.  We find that the early-type components
of pairs with diffuse hot gas appear to be relaxed objects while those
in RR~210 and RR~216, where no diffuse emission has been found,
display unambiguous signatures of ongoing interaction.

New wide-field V and R-band observations are used to study the photometric
properties of the early-type components and to search for {\it candidate} faint
galaxy populations around each of the pairs. While no diffuse optical light is
found for any of the pairs, all of the early-type members show very extended and
concentric luminous envelopes. We identify a faint galaxy sample in each field
and we consider whether they could be physically associated with the luminous pairs based upon $(V-R)$ colors and photometric properties.
We find that the distribution of  $r_e$ and $M_R$ for the {\it candidates}
are similar in three of the fields (RR~143, 216 and 242).  The same selection
criteria applied to the field of RR~210 suggest a fainter and more compact
population possibly suggesting a larger background fraction than in the other
fields.

\end{abstract}


\section{Introduction}

High-redshift studies reveal the existence of a class of objects
with structural, dynamical and stellar population properties very
similar to those of normal nearby early-type galaxies (ETGs)
\citep[(see e.g.][]{Franx03, Chapman04,Treu05}. If ETGs are indeed
the most evolved systems then these high redshift examples
presumably suggest a quasi-primordial nature for ETGs implicit in
the tightness of the fundamental scaling relations. At the same
time, a large amount of observational work indicates that a
relatively large fraction of ETGs show evidence of interactions,
accretions/mergers and recent star formation episodes. This suggests
that predictions of $\Lambda$CDM hierarchical models for ETG
formation cannot be ruled out \citep[see e.g.][]{DeLucia06} even at
low redshift.

Another open question involving ETGs is connected with the possible
dissimilar evolution of these galaxies in different environments
expected from hierarchical models.  Hierarchical galaxy evolution is
assumed to be ongoing with signatures of galaxy reprocessing
possibly more evident in low density environments (LDEs).  Peculiar
morphologies \citep[see e.g.][]{Reduzzi96, Colbert01, Sulentic06}
and kinematical decoupling \citep[see e.g.][]{Falcon05} are indeed
present in a significant fraction of ETGs in LDEs. Recent studies of
line-strength indices identify signatures of very recent
rejuvenation episodes in ETGs located in LDEs and, although there is
a large spread in age, on average these ETGs are about 1.5-2 Gyr
younger than their cluster counterparts \citep[see
e.g.][]{thom05,Annibali06,Clemens06}.

The evolution of local ETGs in LDEs is also the subject of intense
study in the X-ray domain. Chandra and XMM-Newton observations are
clarifying the X-ray properties of loose groups of galaxies which
are the defining aggregates in LDEs. A hot Intra-Group Medium (IGM)
is detected primarily in loose groups containing an ETG population
\citep[see e.g.][]{M00}. Early work \citep{Z99} suggested that
groups might fall into different classes defined by their X-ray
properties: from groups with a luminous, extended, hot IGM centered
on a giant E to groups with little or no diffuse emission.   Several
examples of these classes can now be found in the literature
\citep[see e.g.][] {Mul03,Tri03,Bel03,Ota04}.  In a hierarchical
evolutionary scenario the final product of a merged group is
expected to be an isolated  luminous elliptical with extended X-ray
halo and a few have been identified \citep[see e.g.][]{Mul99,Vik99,
Jon03,Kho04}. At the same time, recent XMM-Newton observations of
optically selected merger remnants like NGC~3921, NGC~7252 and Arp
227 \citep{Nol04,Rampa06} show  that the hot gas is  X-ray
underluminous compared with mature E galaxies into which these
merger remnants are expected to evolve \citep[see also][]{Sansom00}.

Both optical and X-ray observations offer possible evidence for {\it
secular evolution} in LDEs. Indeed galaxy-galaxy tidal encounters
(and mergers) should be more efficient in groups than in clusters
because the velocity dispersion of a group is similar to that of an
individual galaxy. Our goal is to extend optical and X-ray studies
to physically isolated pairs of galaxies which are the simplest and
most common aggregates in LDEs where we can study the evolutionary
effects of galaxy-galaxy encounters at the ``cellular level". Mixed
E+S binary systems are particularly interesting in the context of an
evolutionary accretion scenario \citep[see e.g.][]{RS92,HT01,Dom05}.
Optical and MIR studies provide evidence that the gas-rich spiral
component may cross-fuel the elliptical component \citep{Dom2003} in
some cases indicating a special mechanism for ETG secular evolution.
Study of such relatively simple structures may shed light on a {\it
possible evolutionary link} between poor groups and isolated Es. We
have presently only very limited information on the X-ray properties
of mixed pairs \citep{RR98,HC99,TR01} with detected ETG members
showing X-ray luminosities and $\rm L_X / L_B $ ratios that span the
full range observed for early-type galaxies in general. A group-like
extension has been detected in some pairs but there is no evidence
that pair membership affects the global X-ray properties of galaxies
\citep{TR01}. Detection of diffuse X-ray emission in some pairs
might offer the best means to identify systems that are actively
involved in a hierarchical merging process.

In the above context we present: 1) new XMM-Newton observations of
two E+S pairs (RR~143 and RR~242) both of which show diffuse
emission (plus pairs RR~210 and RR~216 previously observed with
ROSAT \citep{TR01}) and 2) results of an optical photometric search
for tidal features/optical diffuse light and a study of {\it
candidate} faint galaxy populations that might be  associated with
the pairs. Groups studied by \citet{ZM98} revealed a significantly
higher number of faint galaxies ($\approx$50 members down to
magnitudes as faint as M$_B \approx$ -14 +5log$_{10}$ h$_{100}$) in
groups with a hot IGM compared to groups without this component.
Section~2 of this paper reviews the presently known properties of
our sample of pairs. Sections~3 and 4 describe the X-ray and optical
observations as well as data reduction techniques. Individual
results are presented in Section~5 while Section~6 presents an
analysis of the properties of candidate faint galaxy members. A
discussion of the results in the light of current literature is
given in Section~7. We use H$_0$=75 km~s$^{-1}$~Mpc$^{-1}$
throughout the paper.

\section{The picture  so far}

The E+S pairs in our sample were taken from the \citet{ReRa95}
catalogue of isolated pairs in the Southern Hemisphere.
Table~\ref{table1} summarizes photometric, structural and
kinematical properties of pair members extracted from the
literature.

{\bf RR~143} (NGC~2305 + NGC~2307) involves a very isolated pair
with projected component separation R $\approx$ 51 kpc and velocity
difference $\Delta$V $\approx$270 km~s$^{-1}$.  The much larger
velocity difference implied by NED derives from an older and less
accurate measurement for NGC~2307 (see Table~\ref{table1}).  A study
of the ellipticity profile for NGC~2305 (\citet{RR96}) shows that
the outer regions of the galaxy (r$\geq$ 20\arcsec) are on average
flatter ($\epsilon$ $\approx$ 0.3) than the inner ones ($\epsilon$
$\approx$ 0.22-025) consistent with the presence of sub-structure.
Isophotal shape analysis also supports the presence of sub-structure
being boxy within 20\arcsec\ of the center and disky between 20 and
35\arcsec. A recent bulge/disk decomposition study \citep{DeSou04}
suggests that the galaxy is lenticular rather than an elliptical.
Kinematical studies however show that the galaxy has a maximum
rotation velocity $\leq$50 km~s$^{-1}$ \citep{R88} which is more
typical of an E than an S0.

{\bf RR~210}  (NGC~4105 + NGC~4106) form a strongly interacting pair
\citep{AM87} separated by projected 7.7 kpc and 258 km~s$^{-1}$. It can also be described as a loose triplet
\citep{Tully88} if one includes IC~764 with a similar recession
velocity (2127  km~s$^{-1}$) and a  projected separation of
46.3\arcmin\ ($\approx$ 363 kpc). \citet{RR96} obtained B, V and R
surface photometry for  both components which show almost constant
color profiles $(B-V) \approx$0.9-1 suggesting that both are
early-type galaxies. NGC~4105 shows no evidence of perturbation
while NGC~4106 is strongly perturbed with arms/tails that are likely
a product of interaction with the companion. NGC~4105 shows evidence
for a dusty center and disky outer structure. NGC~4106 shows a disky
central structure.

\citet{Long98a,Long98b} obtained kinematic and spectro-photometric
properties for both pair members. Evidence for strong interaction
also comes from velocity and velocity dispersion profiles along a
line connecting the nuclei of the two galaxies. Study of
line-strength indices \citep{Long99} suggests the presence of very
recent star formation episodes. In particular NGC~4106 shows a value
for the H+K(CaII) index larger than 1.3, i.e. larger than the
maximum value attainable in post-starburst models (both with solar
and super-solar metallicity) which suggests the presence of
H$\varepsilon$ in emission. This feature is considered a good
indicator of recent star formation \citep{Rose84,Rose85}. The
detection of [OII]$\lambda$3727-9\AA\ emission in the nucleus of
NGC~4105 is also suggestive of a recent star formation episode
\citep{Long98a}.

\citet{Caon00} showed the extended nature of the  ionized gas
distribution while mapping H$\alpha$ kinematics in NGC~4105. The
ionized gas counterrotates relative to the stars and shows a
different behaviour on opposite sides of the slit (P.A. 151$^\circ$)
centred on the nucleus: the NW side shows a less steep central
gradient, and a dip at about 13\arcsec. This data points toward a
past cross-fuelling or accretion event. The stellar velocity
dispersion is quite high in both galaxies (see Table~\ref{table1}).
NGC~4105 shows slow stellar rotation \citep[(36$\pm$7
km~s$^{-1}$)][]{Caon00} in contrast to the ionized gas rotation
which reaches 240$\pm$20 km~s$^{-1}$. \citet{Long98b} report a
higher stellar rotation velocity $\approx$ 116 km~s$^{-1}$ for
NGC~4106. Modeling of kinematics and surface photometry in NGC~4105
\citep{Samurovic2005} do not support the presence of significant
dark matter at least inside $\approx$1 r$_e$. Their models neglect
any interaction with NGC~4106.

{\bf RR~216} (IC~3290 and NGC~4373) form an isolated pair according to both
\citet{Sadler84} and \citet{AM87} with a projected separation of 26 kpc and
$\Delta$ V = 54 km~s$^{-1}$. The system is in the Hydra-Centaurus region but
located on the outskirts at 4.9$^\circ$ from the cluster center
\citep{Dick86}. B, V, and R surface photometry \citep{RR96} shows
$(B-V)\approx$1 throughout both galaxies although IC~3290 shows a well
developed bar and a ``grand design'' spiral pattern. NGC~4373 shows disky
structure in the isophotal shape profile but the luminosity profile is well
represented by a r$^{1/4}$ law suggesting that the galaxy is a disky E.

{\bf RR~242} (NGC~5090 and NGC~5091) has parameters similar to the other pairs
(R $=17$ kpc, $\Delta$ V = 108 km~s$^{-1}$), but it is found in a richer local
environment than the previous systems. Available NED data suggest that the
pair is part of a loose group. Four additional luminous galaxies with
comparable redshifts can be found within a radius of 300 kpc around NGC~4105:
NGC~5082 (at 5.8\arcmin\ /77.5 kpc, $\Delta$ V$_{pair}$= 421 km~s$^{-1}$),
NGC~5090A (at 20.3\arcmin\ /298.4 kpc, $\Delta$ V$_{pair}$= 45 km~s$^{-1}$),
NGC~5090B (at 13.8\arcmin\ /184.1 kpc, $\Delta$ V$_{pair}$= 773 km~s$^{-1}$)
and ESO~270~G007 (at 24.4\arcmin\ /326.1 kpc, $\Delta$ V$_{pair}$= 275
km~s$^{-1}$). If all above galaxies belong to a group, then it has an average
recession velocity of 3681 km~s$^{-1}$ and a velocity dispersion of 327
km~s$^{-1}$.

NGC~5090 appears to be a``bona fide'' elliptical without signatures
of interaction according to the surface photometry of
\citet{Govo00}. The galaxy hosts an FRI radio source (PKS B1318-434)
with two jets aligned perpendicular to a line connecting the galaxy
nuclei \citep[see e.g.][and references therein]{Lloyd96}.
\citet{Carol93} have obtained the velocity dispersion and rotation
velocity profiles of NGC~5090 which show high central velocity
dispersion and low rotation velocity, both characteristic of E
galaxies. \citet{Bettoni03} estimate a BH mass of 1.1$\times$10$^9$
M$_\odot$ for NGC~5090.

 The X-ray picture given by $ROSAT$ \citep{TR01} indicates that
X-ray luminosities and $L_{\rm X} / L_B$ ratios of pairs encompass a very
wide range, in spite of the very small number of objects studied. In
particular, the 4 systems presented here have very different luminosities
 and X-ray morphologies in contrast to their relatively similar optical
ones (see Table~1). In RR~143 and RR~242 the X-ray emission is much more
extended than the optical light. It can be reasonably interpreted as
associated with a ``group-type" potential. Relatively faint and compact
emission (i.e. within the optical galaxies) is detected in RR~210 and
RR~216. Although no spectral properties could be derived from the
ROSAT-HRI data, the observed luminosities would point to emission from
the galaxy only, possibly from the evolved stellar population, with
little/no contribution from a group component.

\section{Observations and Reduction}

\subsection{X-ray XMM-Newton observations}

We obtained XMM-Newton observations of RR~143 and RR~242 with EPIC
in the Medium filter as reported in Table~\ref{table2}.  We used the
XMM-Newton Science Analysis System (SAS) version
$xmmsas\_20050815\_1803$-6.5.0 to clean the data from flaring events
(which reduced the original exposure by $>$ 30\%: see
Table~\ref{table2}) and to filter them with the standard ``FLAGS"
(see the science threads at {\tt http://xmm.vilspa.esa.es}).  We
retain single, double and quadruple events for morphological and MOS
spectral analysis. We consider only single events for spectral
analysis of the PN data. We also made use of both
$CIAO$\footnote[1]{\tt http://cxc.harvard.edu/ciao/} and
$DS9/Funtools$\footnote[2]{\tt http://head-cfa.harvard.edu/RD}.  CCD
gaps are masked out during the spatial analysis. In order to improve
statistics and to avoid having to take into account the different
CCD gap patterns of the two instruments, we summed all EPIC-MOS data
and kept the EPIC-PN data separate. These two datasets will be used
for all except the spectral analysis in this work.

\subsubsection{Source extent}

We smoothed the EPIC-MOS data with an adaptive filter in order to
better visualize the emission. We show 0.5-2.0 keV isointensity
contours superposed on our optical images (see Section~3.2) in
Figure~\ref{figure1}. The extended emission is centered on the
brighter early-type component (NGC~2305 in RR~143 and NGC~5090 in
RR~242) in both pairs. The spiral galaxies (NGC~2307) in RR~143 and
NGC~5082 west of RR~242 also detected as discrete sources. A number
of additional sources are seen in both fields of the pairs. However
in all but one case (Galaxy \# 13627 in RR~242, Table~8. coincident
with the X-ray source \#23 in Table~10), the X-ray sources are most
likely background objects unrelated to the pairs. We present them in
the Appendix and in Tables~9 and 10 for completeness, but we will
not use use them in the present discussion.

We employed azimuthal averaging to assess the extent of the diffuse
emission in different energy bands.  This is only an approximation,
because the emission is not azimuthally symmetric around either
pair. This approximation allows us to measure a total average extent
and to determine the background level. We excluded discrete sources
from the profiles which are all centered on the early-type
components.

Evaluation of the background for extended sources is always a
challenge, as shown by several authors
\citep[see][]{Nevalien0X,Read0X} and notes to the use of $xmmsas$
software.  Background maps can be obtained as a byproduct of the
detection procedure in the xmmsas software.   A central extended
source is not excised, however, since the program is designed to
also provide a correct background for sources embedded in diffuse
emission. Blank sky fields are available at {\tt
http://xmm.vilspa.esa.es/} and can be used to determine the
background. They are unfortunately based on thin filter data while
ours are medium filter observations.  We therefore analyzed several
images, both from the archive and from our own data, to evaluate the
shape of the background with the medium filter.  A comparison with
the thin filter data indicates no significant differences, except
for a different relative normalization for profiles obtained in
different energy bands.  We have therefore evaluated the background
from the deep field blank sky data, and normalized them to our own
fields in the outer regions, where the radial profiles show an
almost constant surface brightness, very similar in shape to that of
the background fields. The results for RR~143 are given in
Figure~\ref{figure2} as an example. The same figure also shows that
the emission is prominent in the 0.5-2.0 keV profile and becomes
negligible above 3.0 keV.  Since the contribution in the 2-3 keV
range is not significant, and it is confined to the inner regions,
we have used the 0.5-2.0 keV data to study the radial distribution
of the emission.  The same considerations apply to RR~242.

Figures \ref{figure3} and \ref{figure4} show the net, azimuthally
averaged profiles from PN and MOS data separately.  Emission in
RR~143 extends to almost 500$''$ radius ($\sim$ 120 kpc).  The
profile of RR~242 shows a very prominent central source that
dominates emission in the inner $\sim 1'$ region. A  more extended
component extends to r$\sim 700''$ (160 kpc). X-ray contours in
RR~143 show that the emission is far from azimuthally symmetric. We
produced radial profiles at different position angles in order to
quantify the degree of asymmetry. We find that the asymmetry is also
a function of radial distance from the center. This motivated us to
divid the emission into different azimuthal and radial zones
(Figure~\ref{figure3} central and right panels). Figure 3 confirms
significant azimuthal differences in the data which confirm  visual
impressions taken from the iso-intensity contours. Asymmetries are
also visible in the  X-ray emission from NGC~5082 which is  an
SB(r)0 member of the RR~242 system. In Figure~\ref{figure5} we show
the comparison of radial profiles obtained in different quadrants to
illustrate the evidence for azimuthal asymmetry. The photon
distribution is clearly compressed to the SE, while the NW quadrant
appears more extended.

\subsubsection{Spectral analysis}

In order to measure the spectral properties of the X-ray emission we
extracted photons from different regions, including background
regions appropriate for the emission region we want to investigate.
Given the pattern in the PN detectors, regions in MOS and PN are
often not the same. Thew central region is typically the same
because it fits on the same CCD on PN. The more extended emission is
taken from MOS while PN is limited to the region that fits on a
single CCD.  For this reason the relative normalization between
instruments is left free to vary in order to compensate for the
different regions considered.

We used a local determination of the background, when possible, from
an annulus around the region considered (in general this is true for
the MOS1 and MOS2 data sets, and for the central source in the PN
data). This also implies that the surrounding ``galaxy" is used as
background for the central source in RR~242.  For PN, the background
is taken from a region free of source emission and also free of gaps
close to the region of interest.  Since the emission from the source
does not fill the whole field of view in either galaxy, and the
background level where the source ends is not significantly reduced
relative to the central parts we have used a local determination
also for the whole source; this reduces uncertainties due to the
background variations between different observations, unavoidable
when using a different set of data (like the blank fields, that we
cannot use in any case due to the different filters).

The data are binned to increase the statistics.  We have considered binning to
a specific number of photons, which is reasonable for some regions.  However,
since the contribution of the background is not uniform with energy, and is
larger at high energies, we have also considered larger bins mostly at high
energies to increase the signal-to-noise of the net emission in each bin.  In
all cases we have considered only bins that have a S/N of at least 2 $\sigma$
after background subtraction. The data are fitted in XSPEC with either a
plasma (namely MEKAL in the XSPEC acronyms) or a combination of plasma and
bremsstrahlung or power law to account for high energy tails.  Low energy
absorption is also taken into account and abundances are according to
\citet{wilm}. The results are given in Table~\ref{table3} and discussed later
for each source.

\section{Optical wide-field observations}

\subsection{Observations}

CCD imaging of all 4 galaxy pairs and their surrounding fields was
obtained with the ESO 2.2m telescope using the Wide Field Imager
(WFI). The mosaic is composed of 8 EEV ``{\sc ccd~44}'' type CCDs,
each having $2048 \times 4096$ pixels. The pixel scale is
$0\farcs238$ pixel$^{-1}$ yielding a field-of-view of $34 \times 33$
arcmin$^2$. A single field centered on each galaxy pair was observed
in Service Mode between Nov.~2001 and June 2004. Each field was
observed in the Johnson-Bessell $V$ and $R$ bands (ESO filters
V/89--ESO843 and Rc/162--ESO844) with total exposure times of 3000 s
in both filters (see the Observing Log in Table~\ref{table4}).  The
observing conditions were generally photometric.  Standard stars
from \citet{land92} were also observed each night in one or two
reference CCDs. In order to fill the gaps between the CCDs, as well
as to remove bad columns and cosmic ray hits, multiple 500 s images
were recorded for each field.  To this purpose, the images were
suitably dithered in all cases except one (RR~143).

\subsection{Reduction and calibration}

Reduction of the CCD mosaic data was accomplished with
IRAF\footnote{IRAF is distributed by the National Optical Astronomy
Observatories, which are operated by the Association of Universities
for Research in Astronomy, Inc., under cooperative agreement with
the National Science Foundation.} using standard procedures. All
multi-extension images were bias-subtracted and divided by twilight
flat fields using the Mosaic reduction package {\tt mscred}
\citep{vald98}. Images were then astrometrically calibrated and
registered onto a common distortion-free coordinate grid using  {\tt
mscred} and the pipeline script package {\tt wfpred} developed by
two of us at Padua Observatory. Astrometric calibration was done
using the USNO Catalogue \citep{zach+00}. The internal
(image-to-image) astrometric precision is on the order 0.1 arcsec
while the systematic error of object coordinates is $\sim 0.2-0.3$
reflecting the intrinsic accuracy of the USNO Catalogue.

Before co-addition all CCD images had their counts rescaled
(``normalized'') to match the photometric zero point of a reference
CCD. We adopted CCD\#2 as our reference frame and used the relative
zero points of the eight CCDs determined by \citet{rizzi-phd}.  All
CCDs, once referred to a common zero point, were combined into a
single $8000 \times 8000$ image in each band/exposure. All exposures
were finally registered and median-combined to create a final image
for each target, which covers an area of approximately $32\arcmin
\times 32\arcmin$.  In these images all CCD blemishes and gaps are
canceled out, except for RR\,143.  Figures~\ref{figure7},
\ref{figure8}, \ref{figure9} and \ref{figure10} show the regions of
the sky covered by the frames of RR~143, RR~210, RR~216 and RR~242,
respectively.

Since the images in each dataset (Observing Block) were taken
consecutively under photometric sky conditions the median-combined
images were calibrated by adopting the average airmass of each
observation (see Table~\ref{table4}).  Photometric calibration was
accomplished separately for each dataset using standard stars
obtained on the same nights in order to monitor the nightly zero
point variation using the reference CCD\#2.  A set of linear
calibration relations was then computed for each night:

\begin{equation}
V  = v\arcmin  +  a_V (V-R)  +  zp_V \\
\end{equation}

\begin{equation}
R  = r\arcmin  +  a_R (V-R)  +  zp_R \\
\end{equation}

\noindent

where $v\arcmin$ and $r\arcmin$ are the instrumental magnitudes
normalized to 1 s exposure and corrected for atmospheric extinction. The
following extinction coefficients were adopted: $k_V = 0.12$, $k_R =
0.09$. The adopted color terms of the calibration were
$a_V = -0.140 \pm 0.013$ and $a_R = -0.004 \pm 0.016$,
while the zero points are those given in the last column of
Table~\ref{table4}.  The zero point uncertainties of the
calibration relation are of the order 0.03 mag in both $V$ and $R$.
These relations were used to calibrate the instrumental magnitudes
in our photometric catalogs of the target fields (see Sect.~3.4).

\subsection{Surface photometry of bright pair members}

Surface photometry of the bright pair members was carried out with
the ellipse fitting routine in the {\tt STSDAS} package of {\tt
IRAF} and with the {\tt GALFIT} package \citep{Peng02}. While the
{\tt ELLIPSE} task computes a Fourier expansion for each successive
isophote \citep{Jedr87}, resulting in the surface photometric
profiles shown in Figure \ref{figure11}, {\tt GALFIT} was used to
perform a bulge-disk decomposition -- if needed -- and to determine
the parameters of a Sersic model fit to the galaxies bulge
component. The Sersic profile is a generalization of the de
Vaucouleur's law with $\mu(r) \sim r^{1/n}$, where the Sersic
parameter $n$ is a free parameter \citep{Ser68}. This profile is
thus sensitive to structural differences between different kinds of
ETGs providing a better fit to real galaxy profiles.

Two methods were used ro search for fine structure and signatures of
interaction. 1) A model of the  ETG in each pair was derived from
the {\tt ELLIPSE} output and subtracted from the original image. 2)
The image was convolved with an elliptical Gaussian function
yielding an unsharpened version of the image which was then
subtracted from the original image.

\subsection{Detection of diffuse light around bright galaxies}

A search for optical diffuse light around the early-type components
was carried out with the {\tt SourceExtractor} (see also next section).
All sources detected as ``object" were masked yielding a
background-only image where large-scale diffuse light might best be
detected. This procedure is very sensitive to errors in the
background determination which, in our case, are caused by the gaps
between individual CCDs (for RR~143) or the presence of  bright
foreground stars (for RR~242). 

A large extent for all of the systems is measured from our data - a usefull number in this context is the surface brightness of the last detected isophote of the ETG and its semi-major axis. These values (in the R band) are $25.45 \pm 0.01$ mag arcsec$^{-2}$ at 64 kpc (RR~143), $25.48 \pm 0.01$ mag arcsec$^{-2}$ at 38 kpc (RR~210), $26.39 \pm 0.02$ mag arcsec$^{-2}$ at 137 kpc (RR~216) and $24.85 \pm 0.01$ mag arcsec$^{-2}$ at 65 kpc (RR~242).
The typical background fluctuations where used to estimate upper limits above which diffuse light becomes undetectable in our R band images. The
background level was estimated in 10 squares of 20 $\times$ 20 pixel
across the images. This procedure yielded RMS values of $\sim$ 0.3 \% - 1 \% for the different fields. From this values we determined the following limits in mag arcsec$^{-2}$ for the 4 frames: 25.8 for RR~143, 25.4 for RR~210, 26.5 for RR~216 and 25.9 for RR~242.
We found no evidence for extended diffuse light although filamentary structure
was detected in RR~216 (Figure~\ref{figure12}). To quantify the
significance of this detection we compared the typical background
fluctuation in this field (see above) with the excess light of the plume.  The detection limit for this field was determined to be in the order of 0.3 \%, while the intensity of the plume exceeds the mean background value by $\sim$ 3 \% corresponding to a surface brightness of $\sim$ 24 mag/arcsec$^2$.

To compare our results with isolated ETGs we also examined 40\arcmin $\times$ 40\arcmin fields around the 4 ETGs using the POSS II digital R-band images. Low-pass median
filtering of the images revealed in all cases the existence of a
faint, spatially extended envelope which appears roughly symmetrical
around the ETG. The largest envelope was found for the early type
component in  RR~216 with a diameter of about 120~kpc, supporting the above measurements from the new WFI data. 
In this context it is interesting to note that two of the most isolated
ETGs, with luminosity and distance comparable to our pairs, do not
show such extended envelopes \citet{lvm05,Sulentic06}.

\subsection{The candidate faint galaxy population}

\subsubsection{Detection and sample definition}

A population of faint galaxies around each pair was identified from
SExtractor \citep{Bert96} processing of the WFI images. The search
involved a $0.5^\circ \times 0.5^\circ$ field centered on each pair.
We adopted criteria similar to those applied in the search for faint
galaxy populations  in poor groups \citep{Gru05a, Gru05b}. The
algorithm was used to detect extended sources, i.e. sources with a
{\it stellarity} parameter $\leq 0.5$ \citep{Bert96}. The sample was
limited to candidates larger than $a \geq 1.5\arcsec$ and brighter
than $R= 21$ mag.

The expected contribution from background galaxies was estimated
using the program {\tt GalaxyCount}
\footnote{http://www.aao.gov.au/astro/GalaxyCount/}
\citep[see][]{Ellis06}. The program calculates the number of
background galaxies (and standard deviation) expected in a magnitude
limited observation of a given field. This program was used to
emulate the characteristics of our observations and the selection
criteria used for galaxy identification with SExtractor. We adopted
an average seeing $~$0.8\arcsec, a search area A$~$33\arcmin
$\times$ 33\arcmin, and an exposure time t= 50 minutes with the
collecting area of a 2.2m telescope. We further adopt a throughput
of 60\% for the WFI instrument. Using a range of S/N values we
derive for the interval 15.0 $<$ m$_R <$ 19.5 mag
$\approx$170-200$\pm$20 galaxies with a completeness of about 95\%.
The numbers are compatible with the galaxies found in our field of
view suggesting that we are exploring very low galaxy density
regimes. Over-densities, due to galaxy structures, if any, are
expected to be very small.

The color-magnitude diagrams of detected galaxies in the four
respective fields are shown in Figure~\ref{figure13}. Our sample
galaxies follow the well known trend discussed in \citep[see
e.g.][]{Fab73}, where fainter objects have bluer colors, reflecting
the expected lower metallicities of less massive galaxies
\citep{Dre84}. This trend is well defined in the blue color regime
while contamination with background objects becomes evident at the
red end of the distribution. The bold line in Figure~\ref{figure13}
represents the color-magnitude relation for the Virgo cluster
\citep{Vis77} adjusted for the respective pair recession velocities.
A color restriction of $\pm$ 0.5 mag on either side of the relation,
representing generous limits to the early-type sequence, is the most
effective way to remove the majority of background objects
\citep[see e.g.][]{Kho04}. Therefore, all objects with $(V-R) >
1$ were excluded from the sample (dashed line), leading to the
exclusion of 4 - 15 objects/system. The color-restricted candidate
population of all groups then has a mean color of $(V-R) = 0.58 \pm
0.17$ mag.

\subsubsection{Surface photometry of candidate faint members}

Surface photometry was performed on this sample using the {\tt
GALFIT} package. We used an automatic fit with a 1 component Sersic
law ($\mu(r) \sim r^{1/n}$) and obtained parameters such as central
and effective surface brightness ($\mu_0$ and $\mu_e$ respectively),
effective radius $r_e$ and, of course, the Sersic parameter $n$. The
resulting surface photometric properties of all four samples can be
found in Tables~\ref{table5} - \ref{table8}. 
Figure~\ref{figure14} shows some examples of the surface photometry
of faint candidate galaxies where the surface photometric profiles
were obtained with the {\tt IRAF-ELLIPSE} task and are compared to
the {\tt GALFIT} model (solid line in the surface brightness panel).
The central area affected by the seeing (usually $\sim 0.8\arcsec$)
is not shown in the plots. 

\subsubsection{Projected spatial distribution of the faint candidate galaxy population}

The spatial distribution of the faint candidate members was
investigated by computing the local projected density around each
object. This estimate is based on a method described by
\citet{Dre80}. It consists of determining the radius - and the
corresponding area in Mpc$^2$ - within which the 10 nearest
neighbors of each galaxy reside and hence computing the projected
number density of galaxies~Mpc$^{-2}$ for each galaxy position.
These values are then interpolated to a grid using the thin plate
splines interpolation within {\tt EasyMapping}\footnote{{\tt
EasyMapping} is a mapping software developed by Olivier Monnereau
and is available from {\it
http://perso.wanadoo.fr/olivier.monnereau/EasyMapping.htm}}. The
resulting iso-density maps of the color-restricted candidate samples
are shown in Figure \ref{figure16} for each of the four systems of
galaxies.

\section{Results}

\subsection{The X-ray picture}

X-ray maps of RR~143 and RR~242 reveal the presence of a diffuse
intragroup medium in both systems. However, the spatial distribution
of the X-ray photons from the two groups show some differences.
X-ray emission from RR~143 extends out to r $> 100$ kpc and its
morphology is rather disturbed with an anisotropic distribution of
photons that is not an artifact of the smoothing process.
Figure~\ref{figure3} shows that the photon distribution is not the
same at different position angles and radial distances. ROSAT-HRI
data \citep{TR01} already revealed an elongated source associated
with RR~143. One of the directions of elongation is toward the
companion galaxy which could indicate a link/interaction artifact.
However no optical evidence of tidally generated asymmetries has
been detected in either galaxy. The bulk of the emission is centered
on the early-type component although the late-type companion
(NGC~2307) is also detected (source \# 21 in Table~9).  The energy
distribution of photons from this galaxy (namely below and above 2
kev) suggests a mildly absorbed source, which would point toward a
nuclear source as the origin of most of the X-ray emission with
luminosity L$_{(2.0-10.0)}$ $\sim$ 3$\times$ 10$^{40}$ erg~s$^{-1}$.

The spectral characteristics of the extended component can be
modeled with a $\sim 0.53 $keV plasma (0.52-0.56 at 90\% confidence
level, $\chi^2_\nu \sim 1.4$).  The addition of a power law or a
bremsstrahlung model with fixed $\Gamma=1.5$/kT=10 keV gives a
significant improvement ($\Delta \chi^2 > 10$ with one additional
parameter) and accounts for the excess above 2-3 keV.
Figure~\ref{figure6} shows the data and best fit model using the
data from all three instruments. We find no significant differences
if we limit ourselves to the central $20''$ radius or if we consider
a larger area. In all cases we find a feature in the residuals
around 1-1.5 keV that we are not able to model even taking into
account variable abundances, multi-temperatures, and a cooling flow.
All consistently yield temperatures around 0.5 keV (the cooling flow
range is 0.3-0.8) and solar abundances.  We find marginally improved
results with a model where the abundance ratio between different
elements is not fixed: i.e. with Fe overabundant and C
under-abundant relative to the other elements. The residuals at
1-1.5 keV are somewhat smaller but are still present.  Results are
given in Figure~\ref{figure6} for the PN data alone. The absolute
values of the abundances are not constrained so we put little weight
on their relevance in the discussion. The total X-ray luminosity of
the intragroup medium is L$_X$ $\sim$ 3 $\times$ 10$^{41}$ erg
s$^{-1}$ (0.5-2.0 keV) with an average density of n$_e$ $\sim$ 8
$\times$ 10$^{-4}$ in the inner 200\arcsec and a corresponding gas
mass of 8 $\times$ 10$^{9}$ M$_\odot$.

The distribution of X-ray photons in RR~242 is more extended and
less asymmetric than in RR~143.  The source shows a prominent peak
centered on the nucleus of RR~242a (NGC~5090) and consistent with
the position of the radio source PKS B1318 -434 whose axis is
oriented perpendicular to the line joining the pair components. This
suggests that the tidal interaction may be influencing the way that
the central engine of the elliptical is fueled.

The radial profile of the extended emission that is shown in
Figure~\ref{figure4} can be modeled with two, or even three,
components (right panel): at r $<$ 30\arcsec, the emission is
consistent with the instrumental PSF \footnote{We have used the
analytical formula given by the ``EPIC status of calibration and
data analysis" document [XMM-SOC-CAL-TN-0018], of 11/02/05}.  A
significantly flatter component is present at r$>$ 200\arcsec that
can be modeled with a $\beta$-profile with a core radius r$_c$
$\sim$ 420\arcsec\ and $\beta \sim$1 (see figure). These two
components do not completely describe the data: the central region
appears more extended than a simple PSF. The addition of a
$\beta$-profile with r$_c$ $\sim$ 45\arcsec\ and $\beta$ $\sim$1
could account for the excess emission.  This would imply that the
central peak is due to both a nuclear component and a galactic
component.

The spectral results for emission within  a 30\arcsec\ radius are
consistent with the presence of a low luminosity (L$_X$ $\sim$
10$^{41}$ erg s$^{-1}$, 2.0-10 keV) and  mildly absorbed (N$_H$
$\sim$ 10$^{22}$ cm$^{-2}$) AGN surrounded by emission from a
moderately sub-solar 0.61 keV plasma. The luminosity of the latter
component (L$_x$= 5 $\sim$ 10$^{40}$ erg s$^{-1}$) is consistent
with emission from the ISM of the galaxy also taking into account a
possible contribution from the surrounding IGM. The larger scale
component is also consistent with emission from a hot plasma. The
emisison is best modelled using a two-temperature plasma (kT$_1$
$\sim$ 0.3 keV and kT$_2$ $\sim$ 2.5 keV) both with approximately
solar abundance. The total luminosity of the extended component is
L$_X$(0.5-2.0) $\sim$ 3 $\times$ 10$^{41}$ erg s$^{-1}$ and is
dominated by the contribution from the hotter one (ratio about
10:1). The average density in the inner $\sim$50 kpc is n$_e$ $\sim$
2 and 7 $\times$ 10$^{-4}$ cm$^{-3}$ for the softer and harder
components respectively. This decreases to n$_e$ $\sim$ 0.9 and 2
$\times$ 10$^{-4}$ cm$^{-3}$ in the 50-100 kpc region.

The spiral companion is not detected with certainty as a discrete
source; however, a source with ML$\le$ 10 is found coincident with
the position of the galaxy (with f$_{0.5-2.0} \sim 2 \times
10^{-14}$ erg cm$^{-2}$ s$^{-1}$, or L$_x \sim 6 \times 10^{39}$ erg
s$^{-1}$). This would be consistent with the expected luminosity of
a normal spiral galaxy. Figure 1 shows a distortion in the
isointensity contours at the position of the galaxy is visible in
Figure 1 which also suggests some contribution from it. NGC~5082
located W of trhe pair is clearly detected as source \# 8 in
Table~\ref{table242} with a 0.5-2.0 keV luminosity of $\sim$ 8
$\times$ 10$^{40}$ erg s$^{-1}$. The emission in the vicinity of
NGC~5082 appears disturbed with indications of a tail towards the NW
and of compression towards the SE. This can also be seen in Figure
~\ref{figure5} at a distance of 20-50\arcsec (5-13 kpc). This might
indicate transverse motion of this galaxy relative to the group.
Optical data do not show morphological asymmetries that might be a
signature of ongoing interaction.  At the same time the prominent
bar in NGC~5082 could represent the signature of interaction
involving the  group members \citep{No87}.

\subsection{Optical data}

The WFI image of RR~143 is shown in Figure~\ref{figure7} together
with a close-up of the pair after model subtraction of the ETG
designed to reveal possible fine structure. Surface photometry of
the ETG component RR~143a is shown in Figure~\ref{figure11} (top
left). Unfortunately the non-dithered frames for this object
introduces large errors in the Fourier expansion especially from
about 1.5\arcmin\ onwards. The isophotal ellipticity was fixed at
$\epsilon$ = 0.3 in order to extend the fit despite the large area
masked by the gaps. This may lead to an overestimation of the
surface brightness at the outskirts of the galaxy with resultant
overestimation of the structural parameter n. However the surface
brightness profile shows evidence for a flat outer slope indicating
large n. The model obtained with {\tt GALFIT} gives $n$ = 6.2 with
an effective radius of r$_e$ $\sim$ 50\arcsec\ (12 kpc). The
boxiness of the inner isophotes noted by \citet{RR96} can be
confirmed out to $\sim $20\arcsec\ (5 kpc) whereas the shape of the
subsequent isophotes is certainly affected by the presence of a
foreground star $\sim$ 50\arcsec\ from the center of RR~143a. This
star is also responsible for the bump in the $(V-R)$ color profile
as well as for the residuals in the model-subtracted image in
Figure~\ref{figure7}. A shell-like feature can be seen in the
residual image while the unsharpened masked image shows no
signatures of fine structure.  We conclude that apart from a maximum
2\% boxiness of the inner isophotes no fine structure is present in
RR~143a.

Figure~\ref{figure8} shows the interacting pair RR~210 and its
environment. Obvious interaction induced features are seen,
especially in RR~210b. A Gaussian smoothed image reveals tidally
disturbed arms, a possible bar-like structure and fine structure
also in the ETG member RR~210a. The latter feature is probably
caused by a dust-lane: Dust-filaments surround the core of RR~210a
at a radius of $\sim$ 15\arcsec\ (2 kpc) and extend out to $\sim$
30\arcsec\ (4 kpc) on the NW side. Other features in the model
subtracted image are probably due to residuals or may belong to the
highly perturbed RR~210b which can be argued to be in the
foreground. No shell structure is detected in the ETG.

The surface brightness profile for RR~210a is not well fit witha
single component model, most likely due to the presence of RR~210b.
At $\sim$ 40\arcsec\ (6 kpc) light contamination from the neighbor
becomes dominant. We find no evidence for substructure in the
regular extended halo surrounding the pair. Surface photometry of
the elliptical RR~216b (Figure~\ref{figure11}) reveals homogeneous
profiles out to a radius of about 80\arcsec\ (19 kpc) with: a)
smoothly decreasing $\mu$ and $(V-R)$ color, b) slightly increasing
$\epsilon$ and c) constant position angle and Fourier coefficients.
In the galaxy's outer parts ($\sim$ 1.5\arcmin, 20 kpc) the surface
brightness profile becomes very flat. It is best fit with a very
high value of $n$ = 8 and an extended r$_e$ = 150\arcsec\ (35 kpc).
This flattening coincides with strong boxiness of the isophotes
(more than 5\%) and an increased ellipticity. This may argue for the
presence of a second physically distinct component of diffuse
stellar light surrounding the pair. However a large value of $n$ is
also needed to model the central part of RR~216b and the one
component fit is a good representation of the surface brightness
profile over $\sim$8 magnitudes.

Evidence for recent interaction is detected with the model
subtracted image in Figure~\ref{figure9} revealing a ripple east of
the center at a radius of about 80\arcsec\ ($\sim$20 kpc) and an
asymmetric excess of light along the major axis (North). One of the
grand design spiral arms in RR~216a shows a sharp bend towards the
elliptical. The homogeneous color profile throughout the pair (first
reported by \citet{RR96}) argues for RR~216a not being a true spiral
but probably an ETG with tidally excited arms (tails).

The surface brightness profile of the bright elliptical RR~242a is
well represented by a Sersic law with a high $n$ = 7.5 and a very
extended $r_e \sim$ 400'' (98 kpc), a value typical for bright
cluster ellipticals. Residuals of the one component fit point toward
the presence of a second physically distinct component representing
the outer extended halo of the galaxy. Of course the outer slope may
be affected by a bright foreground star that can bias $n$ towards a
higher value. However the central part of RR~242a shows a luminosity
excess, redder color and flattening of the profile towards the
center (coinciding with a change in position angle). This is due to
a circumnuclear disk of absorbing material that is also responsible
for variations in the other profiles in the central part (within
2-3\arcsec\ radius). The visible extension of the disk is $\sim$
2\arcsec\ (0.5 kpc). This nuclear disk has the same position angle
as the spiral RR~242b which shows an asymmetric dust distribution in
its disk with more absorbing material on the side towards RR~242a
(see Figure~\ref{figure10}, {\it bottom left}). We therefore argue
that material transferred (cross-fuelled) from the spiral to the
center of the elliptical is assembled in a central disk and may be
responsible for fueling the AGN in RR~242a. No signature of fine
structure was found. RR~242a appears to be an old and unperturbed
system sharing some photometric characteristics with bright cluster
ellipticals.

Diffuse light was detected only in RR~216 (Figure~\ref{figure12}).
An extended plume emerges towards the north side of the galaxy and
curves towards the east/southeast represents reasonable evidence for
ongoing interaction in this system.  The plume is detected out to a
radius of about 4\arcmin\ (56 kpc) from the center of the ETG member
and spans more than 8\arcmin\ ($\sim$110 kpc) in projection. We
could model the entire surface brightness profile of RR~210a with a
high value of the Sersic parameter $n$ = 6.5 and r$_e$
$\sim$90\arcsec\ (13 kpc).  The model does not match the luminosity
profile at the center (see Figure~\ref{figure11}). Moreover the
outer halo has a slightly different isophotal ellipticity and
position angle relative to the main body of the galaxy.  Although
such extended halos appear to be connected with the presence of
diffuse light \citep[see e.g.][]{Gon05} we did not detect such
emission in our data.

\section{Considerations on the faint candidate members in the four systems}

In this section we present a comparative analysis of the photometric
and structural properties of the faint galaxy populations that we
identified on the basis of $(V-R)$ color. We assume that these
faint objects are at the same distances as the bright systems near
which they are projected. Three of the pairs (but not RR~210) show
very similar recession velocities (Table~\ref{table1}).
Figure~\ref{figure15} shows the distribution of photometric
parameters $r_e$, $n$ and M$_R$ as determined from the fitting
procedure described earlier. We selected 151, 172, 160 and 102
candidate galaxies in the RR~143, RR~210, RR~216 and RR~242 fields
respectively. Figure~\ref{figure16} indicates that they have a
smooth spatial distribution only for the RR~242 field.  Both RR~210
and RR~216 show clumps of denser regions towards the NW of RR~210
and W of RR~216. The galaxy distribution in RR~143 is intermediate
with less prominent concentrations towards the SW and NW.

Figure~\ref{figure15} shows that the Sersic index distribution peaks
at $n \sim$~1 consistent with  the galaxies being either faint
late-type or low-mass early-type dwarf galaxies \citep{Bin84}.  The
$M_R$ distributions of these objects in RR~143, RR~216 and RR~242
are very similar with peaks at $M_{R}\sim -14.8$, $M_{R}\sim -14.3$
and $M_{R}\sim -14.5$ respectively. Figure~\ref{figure15} also shows
that the $r_{e}$ distributions all peak at $r_{e}\sim 0.5$ kpc.

Although the average Sersic index for RR~210 field is $n \sim$ 1,
the distributions of M$_R$ and r$_e$ peak at fainter magnitudes
(M$_R$ $\sim$ -13.5 mag) and at smaller effective radii (r$_e$
$\sim$ 0.3 kpc) with a noticeable lack of objects with r$_e$ between
0.5 and 1 kpc. Furthermore the $M_{R}$ distribution shows a lack of
candidates in the range -15 $\geq M_R \geq$ -18 and an excess
between -12 $\geq M_R \geq$ -14.  The $\sim 2\times$ smaller
distance of this pair could explain the larger number of candidates
with fainter absolute magnitudes but does not account for the
deficiency of objects between $-15 \geq M_R \geq -18$. If we further
consider that the objects appear concentrated in a clump to the
northwest of the pair we expect that most of them are in a separate
structure and most likely a background group.

In order to further investigate the properties of candidate dwarfs
we study the relations between $M_R$, $\mu_0$ and $n$, i.e. between
distance-dependent and distance-independent quantities, shown in
Figure~\ref{figure17}.  It has been suggested \citep{Gra05} that
these quantities should define a linear relation over a range of at
least 10 magnitudes both for giant ellipticals and dwarf galaxies.
We notice that the $M_R - \mu_0$ relation for RR~210 is
systematically offset by 2.3 mag from the relation in the
\citet{Gra05} sample, although it appears to have the same slope.
The magnitude shift corresponds to a shift in distance of a factor
of $\sim 3$, suggesting that most of these objects are background
galaxies which is consistent with the same inference based on their
spatial distribution (see Figure \ref{figure16}). Notice also that
the 2 confirmed members (represented as triangles in the figure) are
close to the value expected from correlation.

In contrast to RR~210, several galaxies in RR~143 and RR~242 lie
above the $M_R - \mu_0$ line. Adopting the slope found by
\citet{Gra05} and fitting the present distribution we obtain a shift
in magnitude of $\sim 1$ for both pairs probably indicating the
presence of {\it residual} background contamination that cannot be
removed without redshift measures. The RR~216 field - which shows no
extended X-ray emission - is intermediate between RR~210 and
RR~143/RR~242.  The projected number density distribution
(Figure~\ref{figure17}) shows some concentrations as well as an
underlying smooth spatial distribution. An interpretation of the
\citet{Gra05} $M_R - \mu_0$ relation, which shows a shift of 1.5
mag, suggests both a larger background contamination with respect to
RR~143 and RR~242 but also a larger fraction of faint galaxy
companions than RR~210.

Analysis of photometric and color properties for faint galaxy
populations around the pairs yields in three cases (RR~143, RR~216
and RR~242) evidence for the presence of an associated dwarf galaxy
population. However, for conclusive evidence spectroscopic data are
required.

\section{Discussion}

A classic problem in the study of galaxy pairs involves the
difficulty to distinguish between bound and unbound systems. The
problem is of great importance since a bound system implies either a
long lasting co-evolution of the galaxies including: 1) matter
accretion events or gas cross-fueling from gas rich spirals to ETGs
\citep{Dom2003} or 2) their final coallescence which, according to
simulations, could lead to the formation of an E-like remnant
\citep[see e.g.][]{Barnes96}. Significant secular evolution of the
original galaxies might also occur during a chance unbound encounter
although processes like cross-fuelling would be disfavored. Bars and
multiple arms are typical structures developed in simulations of
interacting objects and are considered indications of the secular
evolution of these galaxy \citep[see e.g.] [and reference
therein]{No87,salo93}. In the case of bound systems the increased
probability of merging events could link loose groups with,
especially, mixed pairs containing a massive ETG. Such pairs might
be  ``way stations" -- in a hierarchical evolutionary scenario --
toward the complete coalescence of a group.  Several arguments
indicate however that the chemical \citep{tho03}, dynamical
\citep{nip03} and photometric parameters of massive ellipticals are
not compatible with their formation from several major merging
events distributed across a Hubble time \citep{mez03}.

In this context we discuss the picture suggested by photometric
study of the bright members in our E+S pairs. The spiral components
in RR~210 and RR~216 show unambiguous interaction signatures
(Figures~\ref{figure8} and \ref{figure9}). During encounters disc
galaxies are typically more disrupted than the ellipticals because
the latter are hot kinematical systems. NGC~4373 in RR~216 is a
"bona fide" elliptical with an extended envelope indicated by the
large effective radius (35 kpc). The long (110 kpc) and wide (56
kpc) low surface brightness plume northeast of NGC~4105 argues for
an ongoing strong interaction. Simulations \citep[see
e.g.][]{Barnes96} suggest that plume structures are indicative of
{\it very recent} interaction episodes. Arguments have also been
made in support of the hypothesis that the bar and tightly wound
arms in IC~3290 are induced by interaction \citep[see][]{RR96}.
\citet{No87} showed that bars and grand-design spiral arms induced
by an interaction episode form {\it well after the peri-galactic
passage}. We suggest that, in the light of simulations, we may
interpret both features as the results of  {\it long lasting
coevolution} of the RR~216 pair members.

Neither signatures of interaction nor fine structure are found in
the ETG components of RR~143 and RR~242 (NGC~2305 and NGC~5090
respectively) consistent with the same lack of evidence for
perturbation in the spirals.  They appear to be {\it relaxed
systems}. ETGs in RR~210 and RR~216 display signatures of recent
interaction with the presence of fine structure. Both have dust: an
extended dust-lane is observed in NGC~4105. After model subtraction
(see Figure~\ref{figure9}) NGC~4373 shows the presence of ripples
and plumes.  We may characterize the presence of fine structure with
the parameter $\Sigma$ \citep[see e.g.]  [and reference
therein]{Sansom00}. This parameter provides an empirical measure of
the optical disturbance present in a galaxy because its value is
driven by a combination of effects: a) the optical strength of
ripples, b) the number of detected ripples, c) the number of jets,
d) an estimate of boxiness and e) the presence of X-structures.  The
higher the value of $\Sigma$ the higher the morphological
disturbance and the probability that the galaxy is {\it dynamically
young}. Using the standard definition \citep{Sansom00}, we estimate
a value of $\Sigma$=0 for NGC~2305 and NGC~5090 and conservatively
$\Sigma$=1 and $\Sigma$=3.5 for NGC~4105 and NGC~4373 respectively.
The $\Sigma$ values of NGC~2305 and NGC~5090 are both indicative of
their ``relaxed" nature while that for NGC~4105, and in particular
for NGC~4373, a considerably less relaxed state. These values will
be used below together with the X-ray information.

The X-ray picture given by $ROSAT$ observations \citep{TR01}
indicates that luminosities, L$_X$/L$_B$ ratios and morphologies are
very different in the 4 systems. The distribution of L$\rm _X$ for
the four ETGs encompasses nearly the full range for early-type
galaxies. The relatively faint and compact (i.e. within the optical
galaxy) sources detected in RR~210 and RR~216 are consistent with
emission from the evolved stellar population. RR~143 and RR~242 are
more luminous and their emission is extended which is consistent
with the presence of a large amount of hot gas.  The XMM-Newton
observations indicate that this emission encompasses both pair
members (in the case of RR~143) and even another group member
(NGC~5089 in RR~242). This can be interpreted as evidence that the
pairs are bound and embedded in a ``group-like" potential
\citep{M00}.

The X-ray morphologies also suggest that these two groups may be at
different stages of evolution. RR~242 appears more relaxed with a
more regular and azimuthally symmetric emission. The emission in
RR~143, though centered on the early-type galaxy, has a clear
elongation towards the companion (see Figure~\ref{figure3}) possibly
indicating dynamical effects (tidal/ram pressure stripping).  A
similarly shaped halo is observed in a very well studied compact
group, HGC92, where both the optical and the X-ray halo convincingly
suggest that accretion occurs through sequential stripping and that
the smaller size companions contribute substantially to the
characteristics of the diffuse halo.

Figure~\ref{figure18} (left panel) shows the location of our four
ETG pair components in the log~$\rm L_B$ - log~$\rm L_X$ plane. We
have not considered NGC~3605 \citep[see e.g.][]{Eracl01}, NGC~3998
\citep[see e.g.][and reference therein]{Ho97} and NGC~4203
\citep[see e.g.][and reference therein]{Ho01} because their X-ray
emission is dominated by LINER/AGN activity. The \citet{Sansom00}
sample includes galaxies which are dominant members of poor groups
\citep[see e.g.][]{Helsdon01,M93} and these are identified in the
figure. Notice that the X-ray emission is often associated with
groups containing very few galaxies, {\it overlapping the domain of
classical pairs}. NGC~2300 in the figure is the best known example
of this overlap \citep{M93}. This group is dominated by a bright E+S
pair (NGC~2300 and 2276) that is isolated enough to satisfy the
criterion used in compilation of the Catalog of Isolated Pairs
\citep[][KPG~127]{Kara87}. The other members of this loose group
(NGC~2268 and IC~455) are significantly fainter \citep[$\sim $ 2
mag.][]{M93}.

Figure~\ref{figure18} (right panel) plots  L$\rm_X$/L$\rm_B$ values
\citep{Osul01}, for the same set of galaxies, as a function of the
$\Sigma$ parameter. \citet{Sansom00} and \citet{Osul01} discuss the
anti-correlation between L$\rm_X$/L$\rm_B$ and $\Sigma$. The
observation that  only apparently relaxed ETGs are strong X-ray
emitters is argued to be \citep{Sansom00} consistent with the
hypothesis that strong disturbances, such as mergers, cause the
build up of hot gas halos over a time-scale of several gigayears.
\citet{Osul01} further suggest that some of the scatter seen in the
global L$\rm_X$ versus L$\rm_B$ relation is due to the evolutionary
state and past merger history of early-type galaxies.

As expected both ``relaxed" galaxies (NGC~5090 and NGC~2305) as well
as NGC~4373 that displays the clearest evidence of ongoing
interaction are consistent with the L$\rm_X/ L_B$-$\Sigma$ relation
derived from the \citet{Sansom00} sample. Moderately perturbed
NGC~4105 is a very weak X-ray source with a luminosity similar to
NGC~474 the ETG member of Arp 227 \citep{Rampa06} which displays a
complex system of shells, presumably due to strong past
interactions.  \citet{Read98} found that relaxed remnants (about 1.5
Gyr after a merger) appear relatively devoid of hot gas after
showing an increase in the emission from such gas at earlier stages
\citep[see also][]{Hibbard94}.  \citet{Fabbiano95} noted  that
merger remnants are X-ray weak and suggest that a merger-induced
starburst could drive a galactic superwind which clears the remnant
of most hot gas. After the starburst subsides the hot gas is
gradually replenished by stellar mass loss \citep[for a review,
see][]{Sarazin97} and, perhaps to some small extent, also by the
thermalization of returning tidal debris \citep{Hibbard96}.

The above considerations suggest that RR~143 and RR~242 form bound
structures on the basis of their extended X-ray emission.  Note
however that, at odds with RR~242, RR~143 is a very poor structure
without luminous companions.  RR~210 and RR~216 do not have extended
X-ray emission but both have some luminous companions (see
Figures~\ref{figure8} and Figure~\ref{figure9}) and probably
represent the {\it active} part of very poor and loose {\it
evolving} groups. The {\it activity} is seen in the optically
distorted morphology traced by low luminosity plumes. The presence of
these structures, at least in RR~216, could be interpreted in terms of a long lasting coevolution of the pair.

Additional arguments in favour of the bound nature of some groups
have been provided by \citet{ZM98}. They found a significantly
higher number of faint galaxies ($\approx$50 members down to
magnitudes as faint as M$_B \approx$ -14 +5log$_{10}$ h$_{100}$) in
groups with a hot IGM. This finding led to the idea of defining
aggregates in LDE according to their X-ray properties, ETG fraction
and faint galaxy population. In order to clarify the proposed
connection between poor groups and isolated ellipticals we used our
wide field images to tidentify candidate dwarf galaxy populations
around our pairs. We identify a relatively high number of potential
members in three out of four pairs (e.g. in RR~143, RR~216 and
RR~242) while the faint galaxy sample detected in the RR~210 field
does not appear to be associated with the pair.

Both the X-ray luminosities of RR~242 and RR~143, and the presence
of a large {\it candidate} faint companion populations argue in
favor of the hypothesis that these are bound systems. In the
\citet{Z99} scheme they can be considered as {\it evolved systems}
with X-ray emission centered on the bright E. RR~216 is probably an
intermediate example  of an {\it evolving} groups since its
elliptical is quite perturbed with a modest hot IGM component and
possibly large number of faint companions. The X-ray luminosity in
RR~210 does not support any hot IGM component as in the case of Arp
227 \citep{Rampa06}. The suggested lack of a faint galaxy population
suggests that it is either unbound or in a very early-phase of its
evolution.

We would like to add some considerations on the evolution of the
ETGs in these systems.  From the optical point of view, all the
ellipticals in our sample are evolved, luminous and massive
\footnote {The velocity dispersion, $\sigma$, does not have a
one-to-one relationship with the mass of the galaxy. Recent work by
\citet{Cappellari05} has investigated the relation between the
galaxy mass-to-light (M/L) ratio and the line-of-sight component of
the velocity dispersion with the effective radius. They provide
relations (their equations 7 and 10) which allow a transformation
from measured velocity dispersion to galaxy mass;
\begin{equation}
M_{10} = (16.5 \pm 7.8)\;\sigma_{200}^{3.11 \pm 0.43}
\end{equation}
where $M_{10}$ is the galaxy mass in units of $10^{10}\,\rm M_{\odot}$ and
$\sigma_{200}$ is the velocity dispersion in units of $200 \;\rm km\,s^{-1}$.
The errors have been propagated from the parameter values found by
\citet{Cappellari05}.}  , with an average $\langle \sigma \rangle$ =254$\pm$34
km~s$^{-1}$ (i.e. corresponding to M= 3.47$\times$10$^{11}$ M$_\odot$).

Our systems have masses and environmental properties similar to the class of
``fossil groups'' that is now emerging in the literature, $e.g.$ NGC~1132
\citep
{TD81}, NGC~6482 \citep{Faberetal} or ESO~3060170 \citep{Beuing}.
This is of special importance given the recent evidence of
\citep{Clemens06} which finds that the timing of the formation of
early-type galaxies is determined by the environment but the details
of the process of star formation are entirely regulated by the
galaxy mass.  Since these two quantities -- environmental density
and galaxy masses -- are very similar throughout our sample we
suggest that the difference in their X--ray emission could be
explained in terms of the present evolutionary phase of the system
\citep[see e.g.][]{Sansom00,Osul01}.

\section{Summary and conclusions}

We present an optical and X-ray study of four E+S pairs, RR~143,
RR~210, RR~216 and RR~242 and their  environments with the main aim
to assess their evolutionary histories. We have obtained new
XMM-Newton data for  2 pairs, RR~143 and RR~242, that complement the
ROSAT data already presented the other two
\citet{TR01}.   With these data:\\
$\bullet$ We confirm the presence of extended X-ray emission in both
RR~143 and RR~242 with luminosities
L$_X \sim 3 \times 10^{41}$ [erg~s$^{-1}$].\\
$\bullet$ The emission in RR~143 is centered on the ETG but is
asymmetric and  elongated towards the late type companion with a
total extension to
r $\sim$ 500\arcsec\ (120 kpc). \\
$\bullet$ The large scale emission from RR~242 is more regular and
extends out to 700\arcsec\ (160 kpc).  A low luminosity (L$_X \sim
10^{41}$ erg s$^{-1}$, 2.0-10 keV), mildly absorbed (N$_H \sim
10^{22}$ cm$^{-2}$) AGN is present at the center of RR~242, together
with a moderately subsolar, $\sim$ 0.6 keV, plasma emission,
possibly due to the central galaxy.

We considered the X-ray properties together with the results of a V,
R photometric study of the brighter members and a search for
{\it candidate} faint companions. \\
$\bullet$ We find that the ETGs in X-ray luminous pairs appear to be
relaxed objects while the X-ray faint ETGs display unambiguous
signatures of ongoing interaction. \\
$\bullet$ We identified a sample of faint galaxies among which we may find objects possibly associated with our pairs. We considered the likelihood of physical association with the pairs and find that a small fraction of the detected objects are likely faint companions except in the case of RR~210 where even a larger background contamination is suspected.

In the light of current ideas we suggest that:\\
$\bullet$ Both X-ray luminous E+S pairs RR~143 and RR~242 are likely
the dominant members of {\it evolved} poor groups. Only a few
luminous members are associated with these systems although our
study suggests that they might have a significant population of
faint
member. \\
$\bullet$ RR~210 and RR~216 are likely the dominant members of {\it
evolving} groups.
\\ $\bullet$ At odds with their optical, kinematic (mass) and
environmental similarities, the ETGs in our E+S pairs have
remarkably different X-ray properties. We suggest that this can be
explained either in terms of the present evolutionary phase of the
system or the past histories of the ETGs.

Spectroscopic observations currently on-going with ESO VLT + VIMOS
will provide redshifts for the faint galaxy populations  that will
identify associated dwarf populations and allow us to better study
the evolutionary stage of these systems.

\acknowledgments

RG and RR thank the INAF-Osservatorio Astronomico di Padova and the Institut
f\"ur Astronomie Universit\"at Wien, respectively, for the kind hospitality
during the paper preparation.  RG, RR and WWZ acknowledge the support of the
Austrian and Italian Foreign Offices in the framework science and technology
bilateral collaboration (project number 25/2004). RG and WWZ acknowledge the
support of the Austrian Science Fund (project P14783).  RR and GT acknowledge
the partial support of the Agenzia Spaziale Italiana under contract ASI-INAF
I/023/05/0.  This research has made use of the NASA/IPAC Extragalactic
Database (NED) which is operated by the Jet Propulsion Laboratory, California
Institute of Technology, under contract with the National Aeronautics and
Space Administration. The Digitized Sky Survey (DSS) was produced at the Space
Telescope Science Institute under U.S. Government grant NAG W-2166. The images
of these surveys are based on photographic data obtained using the Oschin
Schmidt Telescope on Palomar Mountain and the UK Schmidt Telescope. The plates
were processed into the present compressed digital form with the permission of
these institutions.  This research has made use of SAOImage DS9, developed by
Smithsonian Astrophysical Observatory.

Facilities: \facility{XMM-Newton}, \facility{2.2 ESO + WFI}.

\appendix

\section{X-ray Individual sources}

We used standard detection procedure to derive parameters for other
X-ray sources detected in the fields of RR~143 and RR~242. We
considered the 0.5-2.0; 2.0-10.0; and 0.5-10.0 keV bands and all
three detectors. Since the detection of serendipitous sources in the
fields was not s motivation for this work we limit ourselves to
sources above a maximum likelihood threshold of 15 in order to
minimize contamination from spurious sources.  For this reason we
did not consider sources within $200''$ of the center of the
emission (except for \#17, \#18, and \#20 in RR~143 clearly visible
on the images) where the algorithm finds a large number of sources
likely to be fluctuations within the extended emission. Fluxes are
determined assuming a power law spectrum with $\Gamma$=1.7 and are
corrected for line-of-sight absorption. When possible we use the PN
data to determine count rates and fluxes, however when the sources
fall ont a CCD gap or defect we use the MOS1 data (see last column
in the tables). Detected sources are listed in Tables 9 and 10.

Several of the X-ray sources coincide with faint optical counterparts.
However, comparison between the candidate member lists with the X-ray
source lists indicate coincidence only between galaxy \# 13627 in  RR~242,
and source \#23.  This is not surprising, since the galaxies are all
significantly fainter than the pair members and are not expected to be
detectable, unless they contain an active nucleus.  Most other X-ray
sources are likely to be background AGNs.

%


\begin{figure*}
\epsscale{1}
\caption{ Iso-intensity contours of the adaptively smoothed X-ray
emission from RR~143 (NGC~2305/ NGC~2307; left) and RR~242
(NGC~5090/NGC~5091; right) detected with XMM-Newton. Contours
are  superposed on R-band images obtained with the 2.2m MPG-ESO
telescope (see text).  The galaxy west of NGC~5090 is NGC~5082.
\label{figure1}}
\end{figure*}

\clearpage


\begin{figure}
\epsscale{1}
\plotone{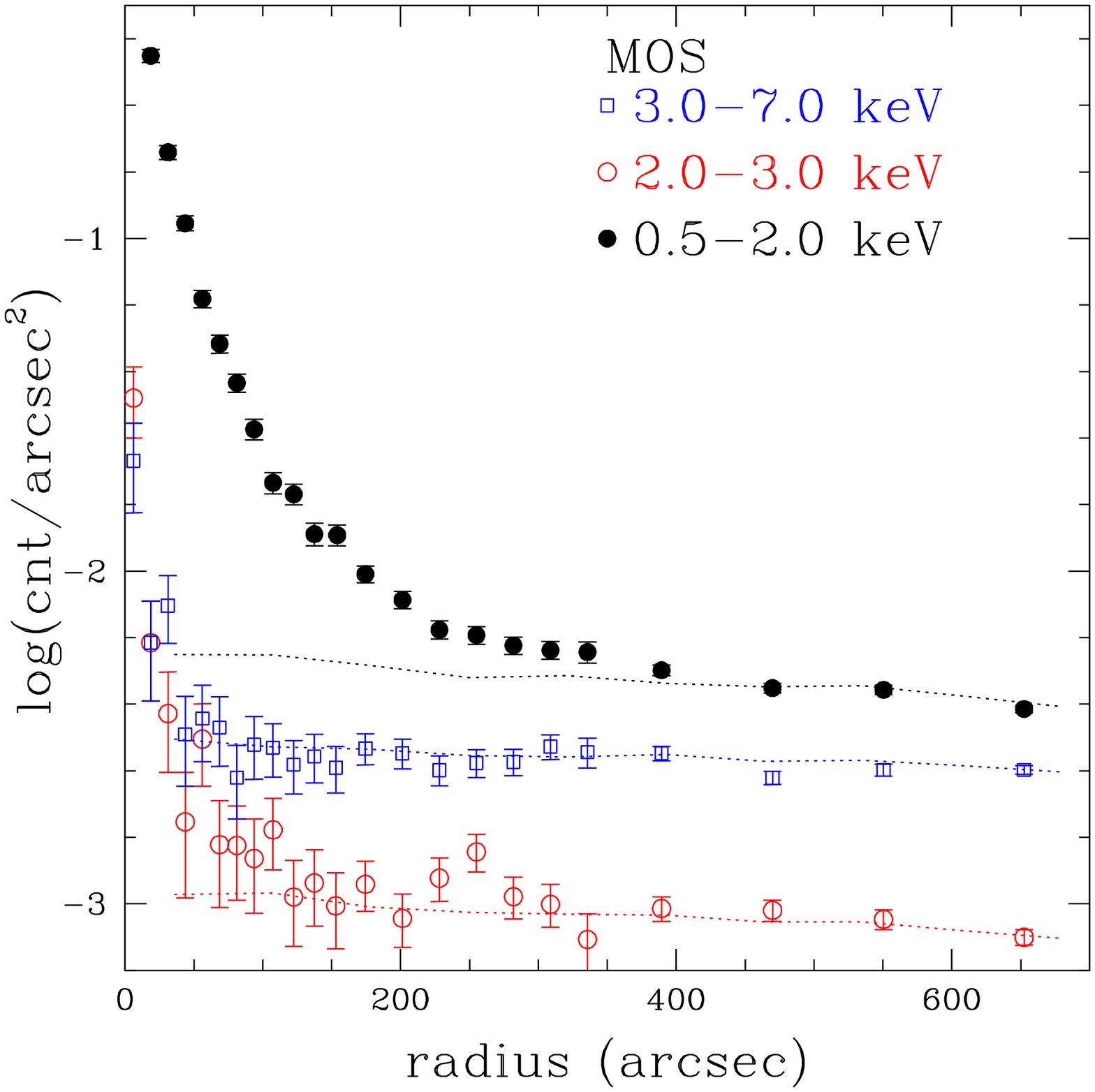}
\plotone{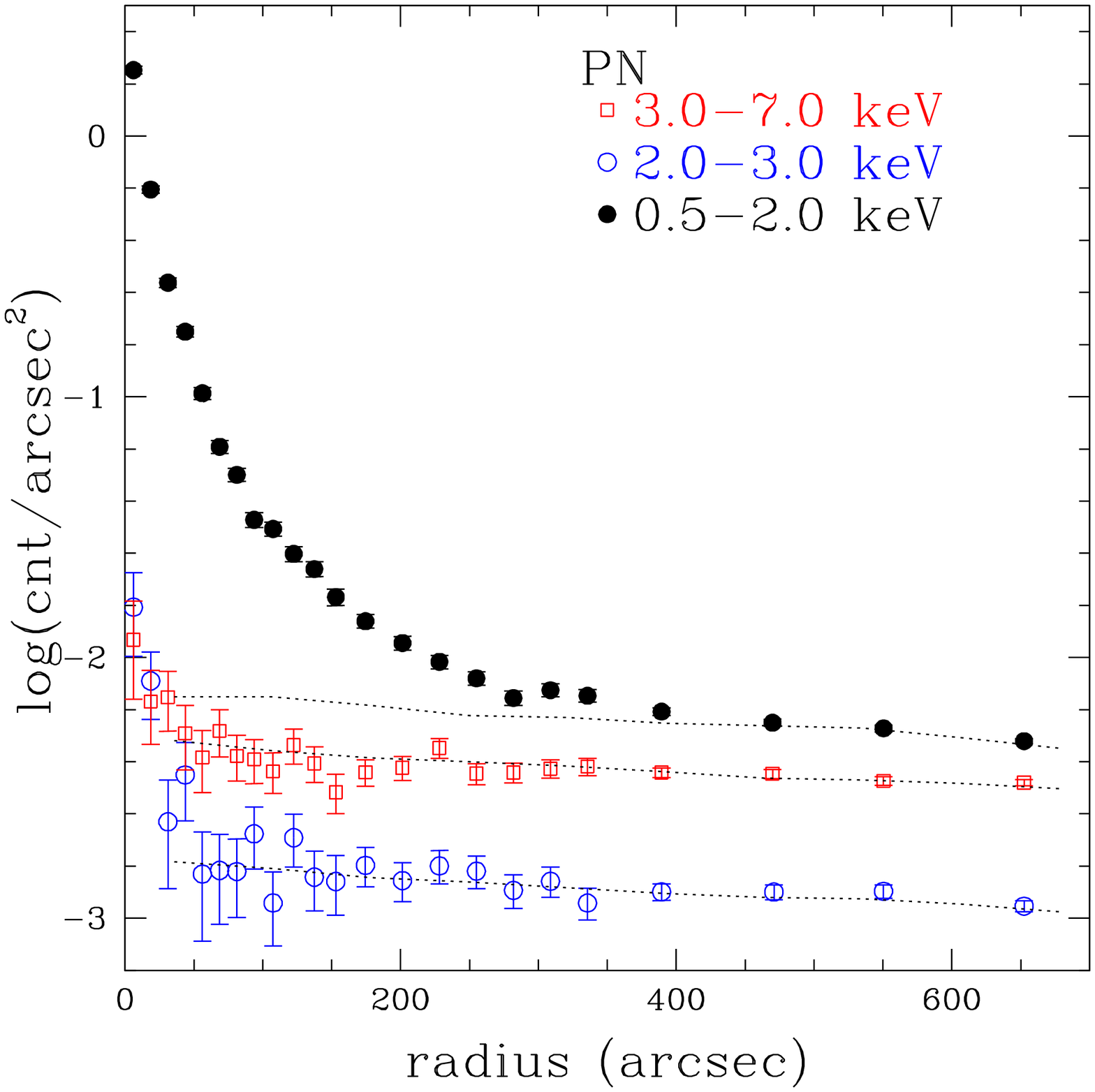}
\caption{ Surface brightness
profile of the raw XMM data in different energy bands, centered on
NGC~2305, the early-type member of the pair RR~143. MOS and PN are shown
separately. The background shapes relative to each profile are also
plotted. \label{figure2}}
\end{figure}


\begin{figure*}
\resizebox{16.7cm}{!}{
\epsscale{1}
\plotone{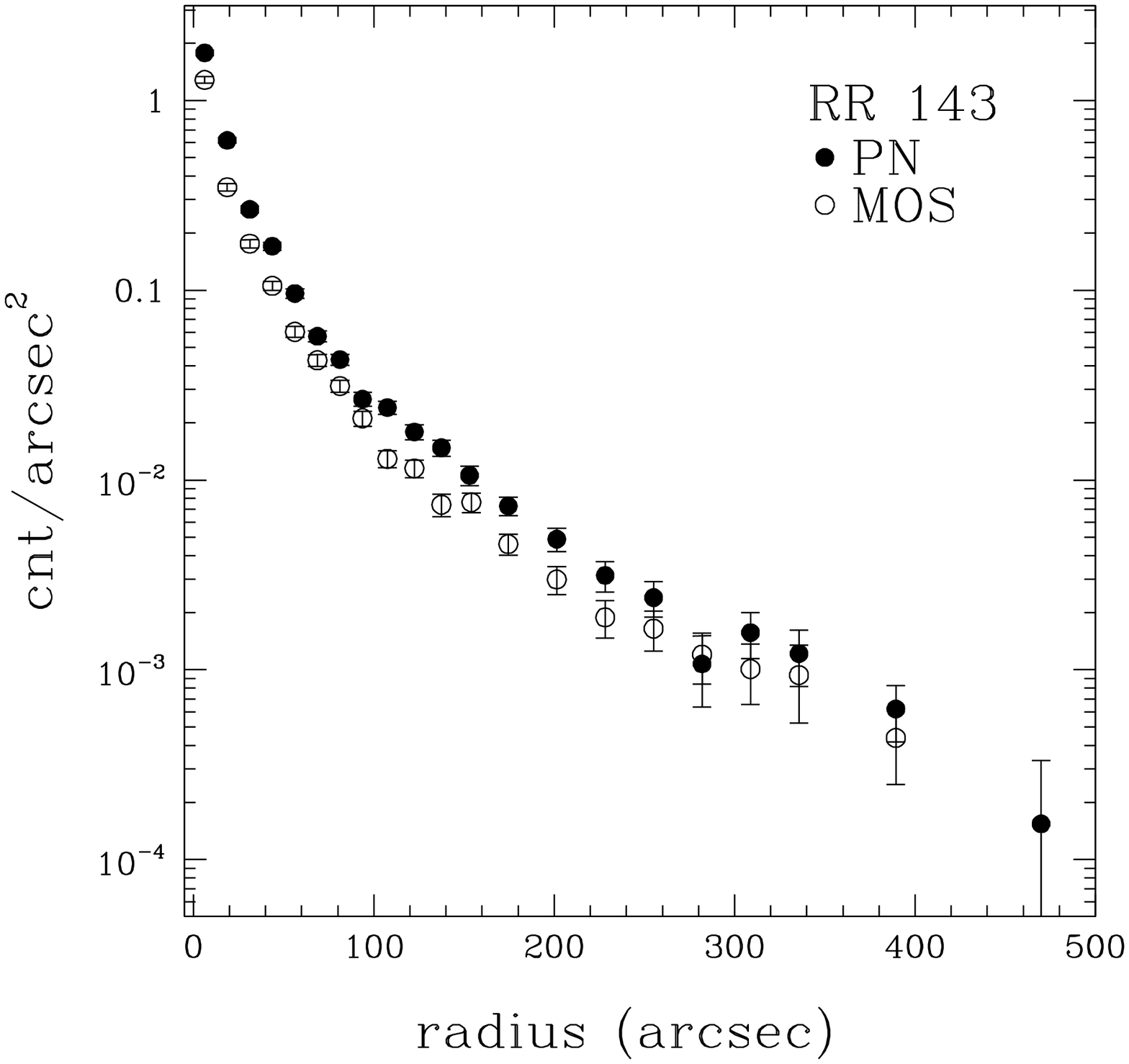}
\epsscale{1}
\plotone{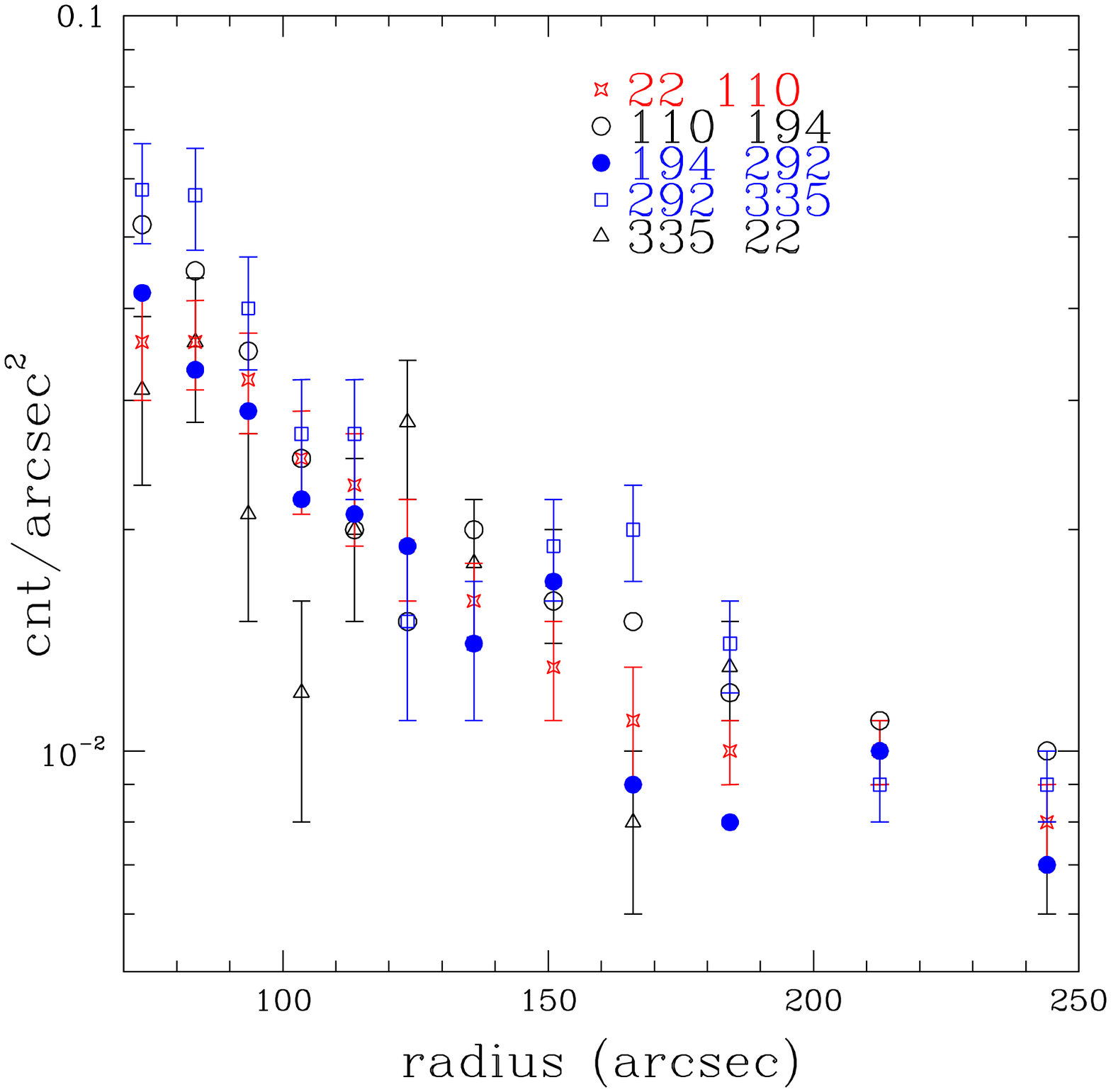}
\plotone{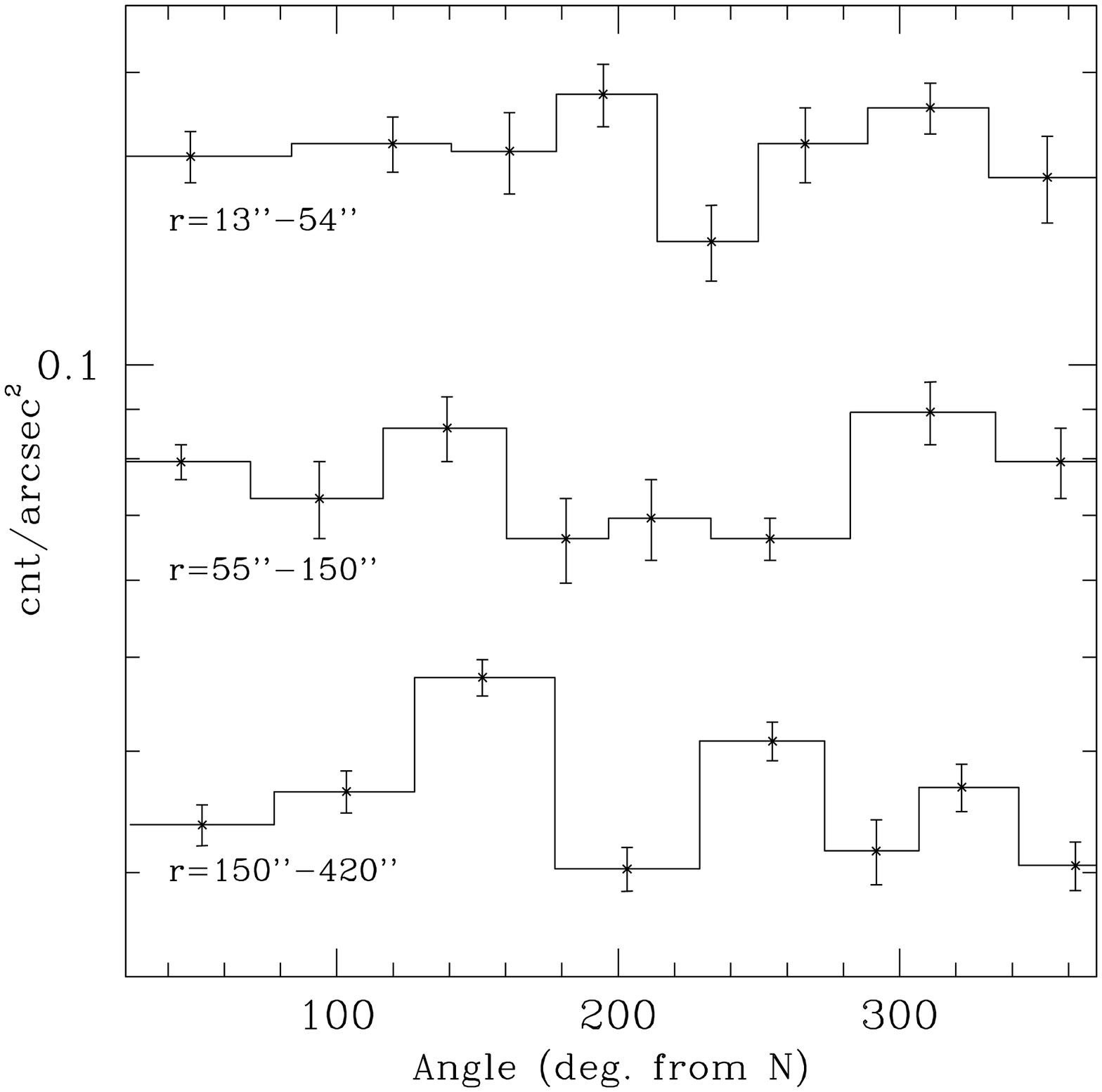}
}
\caption{  Surface brightness profile of the net emission in RR~143:
azimuthally averaged from MOS and PN separately (left), in different
quadrants (middle) and at different angles and radial distance from the
center (right).  All plots are in the 0.5-2.0 keV band.
\label{figure3}}
\end{figure*}

\clearpage



\begin{figure}
\epsscale{1}
\plotone{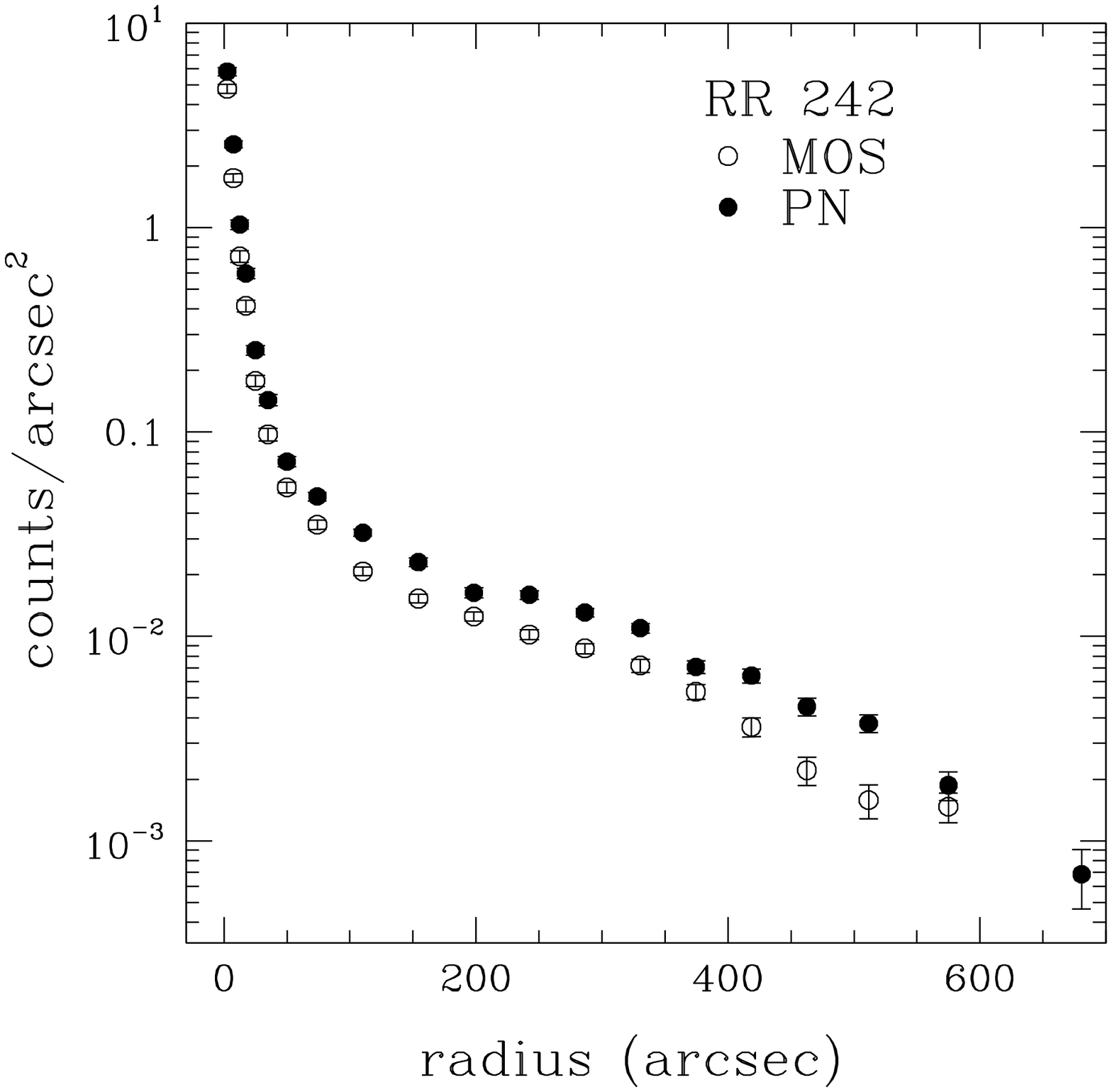}
\plotone{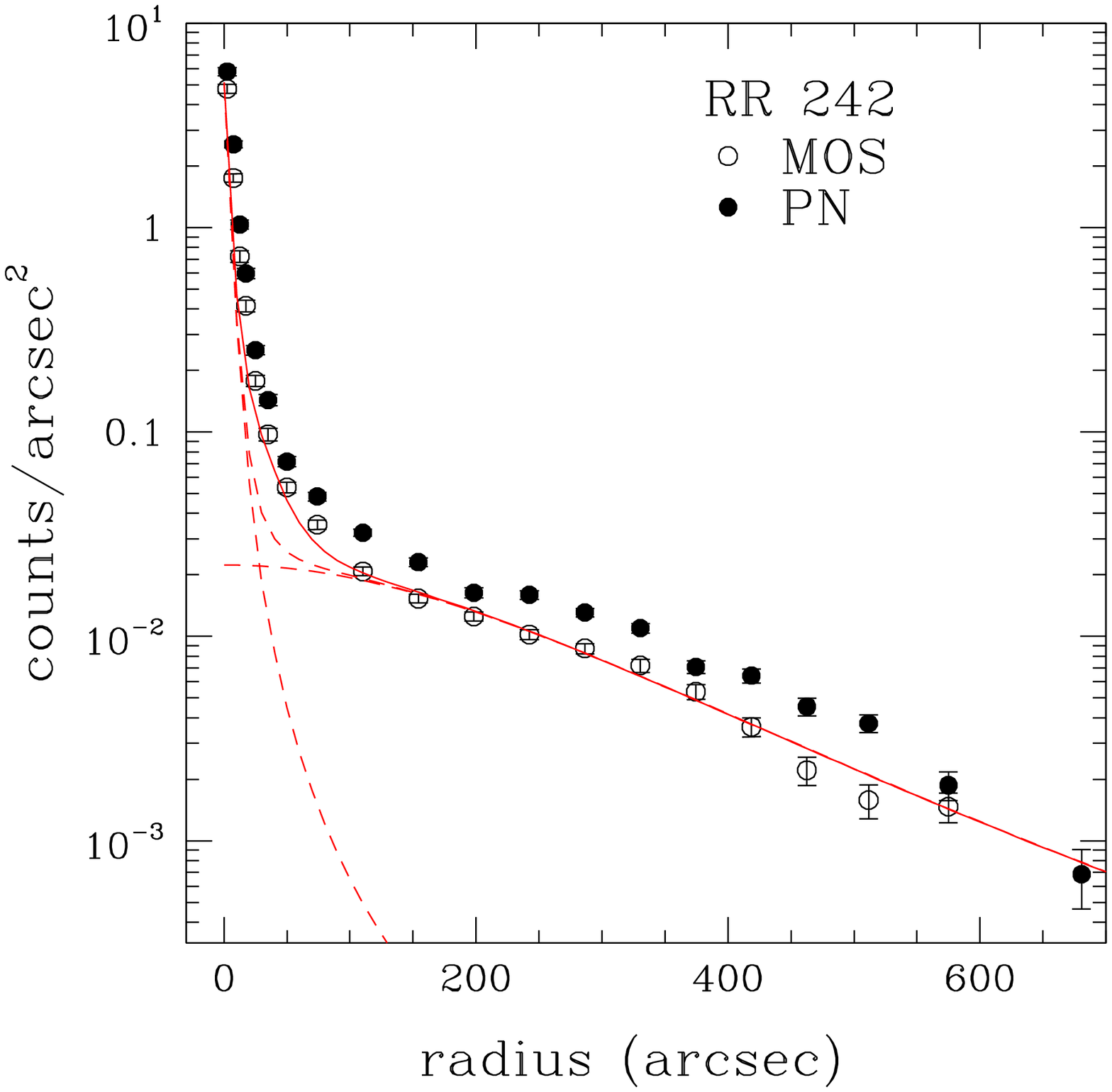}

\caption{Surface brightness profile of the azimuthally averaged
net emission in RR~242, in the 0.5-2.0 keV band, from MOS and PN
separately (left).  A three component model, due to the sum of
the MOS point spread function (PSF) and two $\beta$-profile
functions with r$_c=420''$ and r$_c=45''$ and $\beta
\sim 1$, is shown as a solid line. The PSF and the larger
$\beta$-profile are also plotted as a dashed line (right panel).
\label{figure4}}
\end{figure}



\begin{figure}
\epsscale{1}
\plotone{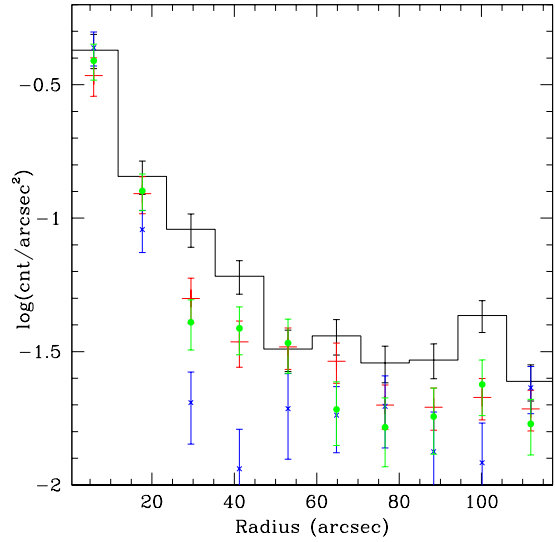}
\caption{Radial profile of the emission centered on NGC~5082 in 4
different azimuthal sectors.  The data are in the 0.5-2 keV energy range.
The histogram refers to the 27-117\degr quadrant, red crosses to
177-207\degr, blue crosses to 207-297\degr and green dots to 297-27\degr 
(clockwise from N).
\label{figure5}}
\end{figure}

\clearpage



\begin{figure*}
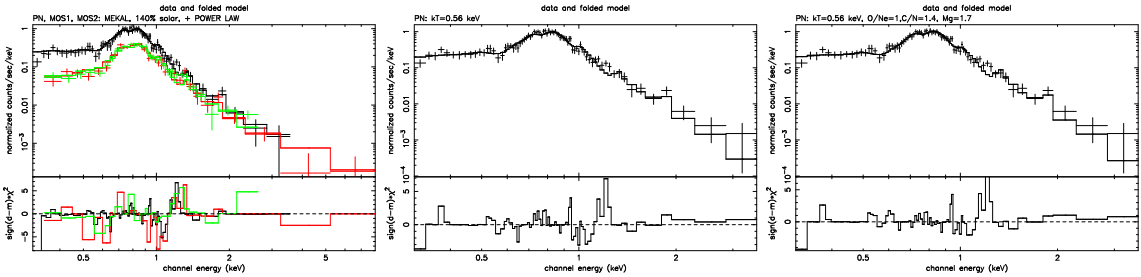

\centerline{
\includegraphics[angle=-90,width=.3\textwidth]{f6a.eps}
\includegraphics[angle=-90,width=.3\textwidth]{f6b.eps}
\includegraphics[angle=-90,width=.3\textwidth]{f6c.eps}
}

\caption{  Spectral distribution of the diffuse emission in RR~143.
Note that since the emission extends over
several CCDs in EPIC-PN, spectral data are obtained from smaller regions
that would fit within a single CCD. To compensate for different areas in EPIC-PN and EPIC-MOS,
the relative normalization is left free.
Top: EPIC-PN and EPIC-MOS data. The model is a
MEKAL (kT= 0.5 keV, 100 \% solar abundances) plus
and power law (fixed $\Gamma=1.5$), with galactic N$_H$.
Middle and bottom:  PN data only.  The model
is a kT= 0.5 keV MEKAL, 90\%  abundances with ratios
between elements  fixed
at the solar values, and a kT= 0.5 keV VMEKAL, with Mg at 170\%, O--Ne
at 100\% and C--N at 140\% of the solar value, respectively.
\label{figure6}}
\end{figure*}

\clearpage



\begin{figure*}
\epsscale{1}
\resizebox{16.7cm}{!}{
\vspace*{2cm} 

}
\caption{
({\it large right panel}) WFI R--band image of RR~143 and its environment.
({\it top left panel}) Finding chart of spectroscopically confirmed members of
the RR~143 system within the FoV.  Galaxies are marked with symbols sized
according to their absolute magnitude. Labels 1 and 2 stand for NGC~2305 and
for NGC~2307 respectively.  ({\it mid left panel}) Residual image of the
central pair after the subtraction of a model of the dominant ETG member;
({\it bottom left panel:}) Residual image after the subtraction of the
gaussian smoothed image.
\label{figure7}}
\end{figure*}



\begin{figure*}
\epsscale{1}
\resizebox{16.7cm}{!}{
}
\caption{({\it large right panel}) WFI R--band image of RR~210 and its environment.
({\it top left panel}) Finding chart of spectroscopically confirmed members of
the RR~210 system within the FoV.  Galaxies are marked with symbols sized
according to their absolute magnitude. NGC~4105 (label 1), NGC~4106 (label 2),
2MASX~J12065029-2936236 (label 3: $V_{hel} = 2019$ km/s),
2MASX~J12063106-2951336 (label 4: $V_{hel} = 2132$ km/s), IC~2996 (label 5:
$V_{hel} = 2256$ km/s).  ({\it mid left panel}) Residual image of the central
pair after the subtraction of a model of the dominant ETG member; ({\it bottom
left panel:}) Residual image after the subtraction of the gaussian smoothed
image.
\label{figure8}}
\end{figure*}



\begin{figure*}
\epsscale{1}
\resizebox{16.7cm}{!}{
}
\caption{({\it large right panel}) WFI R--band image of RR~216 and its environment.
({\it top left panel}) Finding chart of spectroscopically confirmed members of the
RR~216 system within the FoV.
Galaxies are marked with symbols sized according to their absolute
magnitude. NGC~4373 (label 1), IC~3290 (label 2), 6dF~J1225118-393507
(label 3: $V_{hel} = 2140$ km/s), ESO~322-IG~002 (label 4: $V_{hel} = 3264$ km/s).
({\it mid left panel}) Residual image of the central pair after the subtraction of a model
of the dominant ETG member;  ({\it bottom left panel:}) Residual image after
the subtraction of a model of the dominante ETG member; ({\it bottom left
  panel:}) Residual image after the subtraction of the gaussian smoothed image.
\label{figure9}}
\end{figure*}



\begin{figure*}
\epsscale{1}
\resizebox{16.7cm}{!}{
}
\caption{({\it large right panel}) WFI R--band image of RR~242 and its environment.
({\it top left panel}) Finding chart of spectroscopically confirmed members of
the RR~242 system within the FOV.  Galaxies are marked with symbols sized
according to their absolute magnitude. NGC~5090 (label 1), NGC~5091 (label 2),
NGC~5082 (label 3: $V_{hel} = 3896$ km/s), NGC~5090B (label 4: $V_{hel} =
4248$ km/s), ESO~270-~G~001 (label 5: $V_{hel} = 2960$ km/s),
2MASX~J13201668-4327195 (label 6: $V_{hel} = 3088$ km/s).  ({\it mid left
panel}) Residual image of the central pair after the subtraction of a model of
the dominant ETG member; ({\it bottom left panel:}) Residual image after the
subtraction of the gaussian smoothed image.
\label{figure10}}
\end{figure*}


\begin{figure*}
\epsscale{1}
\resizebox{15.7cm}{!}{
}
\caption{  Surface photometry of the elliptical pair member: {\it
from top to bottom} surface brightness, ellipticity, position angle,
fourier coefficients and color profile.
\label{figure11}}
\end{figure*}

\clearpage



\begin{figure*}
\epsscale{1}
\caption{  Detection of diffuse light in RR~216. An extended plume
indicative of on-going strong interaction is visible in the
NE quadrant, with the tip of the plume wrapping towards S.
\label{figure12}}
\end{figure*}



\begin{figure*}
\epsscale{1}
\caption{ $(V-R)$ color-magnitude relation for all pairs/groups. The
bold line is the color magnitude relation for the Virgo cluster
(shifted to the pairs redshift), while the dashed line represents the color restriction at $(V-R) = 1$ applied to exclude background objects. The
spectroscopic confirmed pair/group member galaxies are labeled
as in Figures \ref{figure7} - \ref{figure10}.
\label{figure13}}
\end{figure*}



\begin{figure*}
\resizebox{16.7cm}{!}{
\plottwo{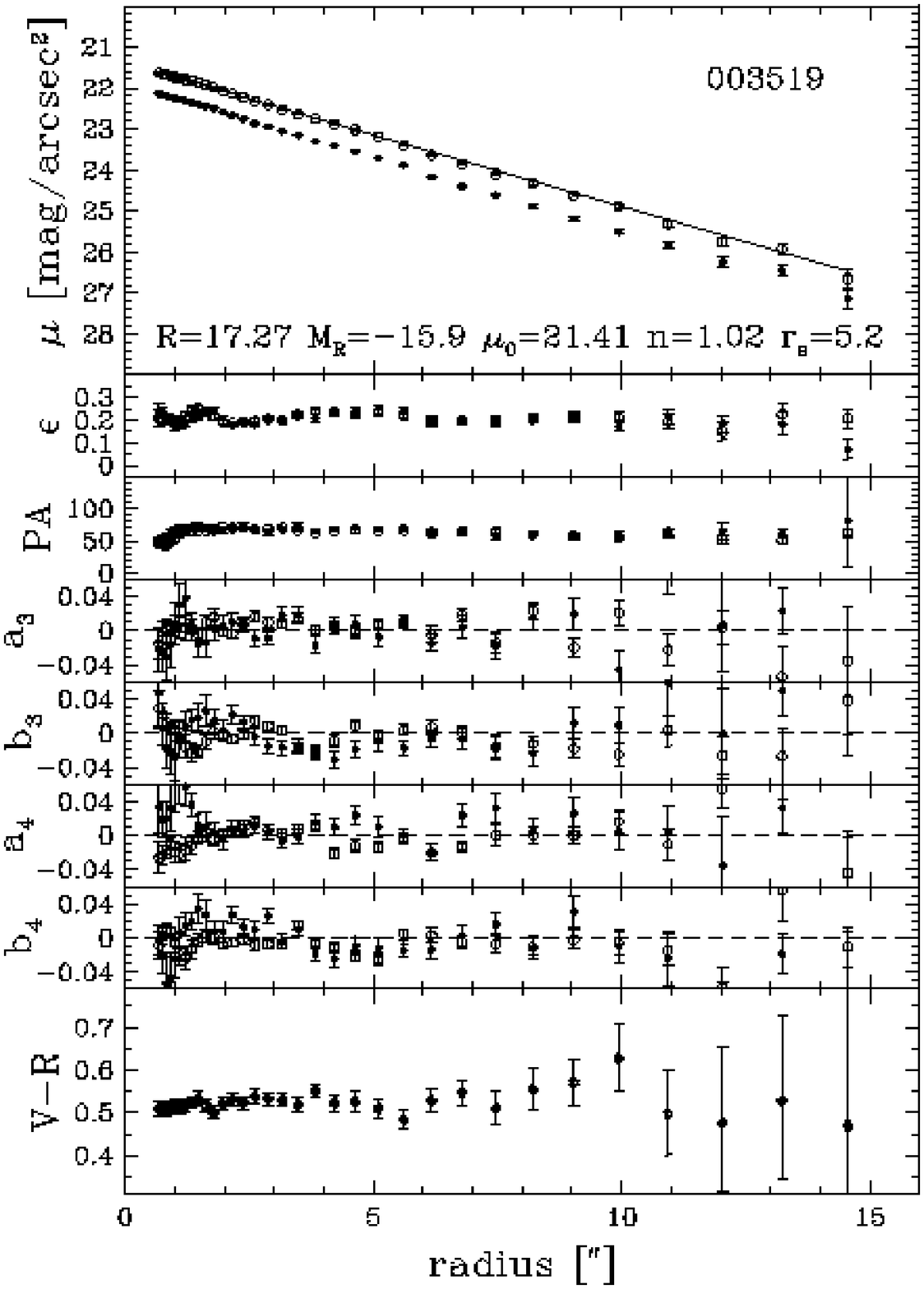}{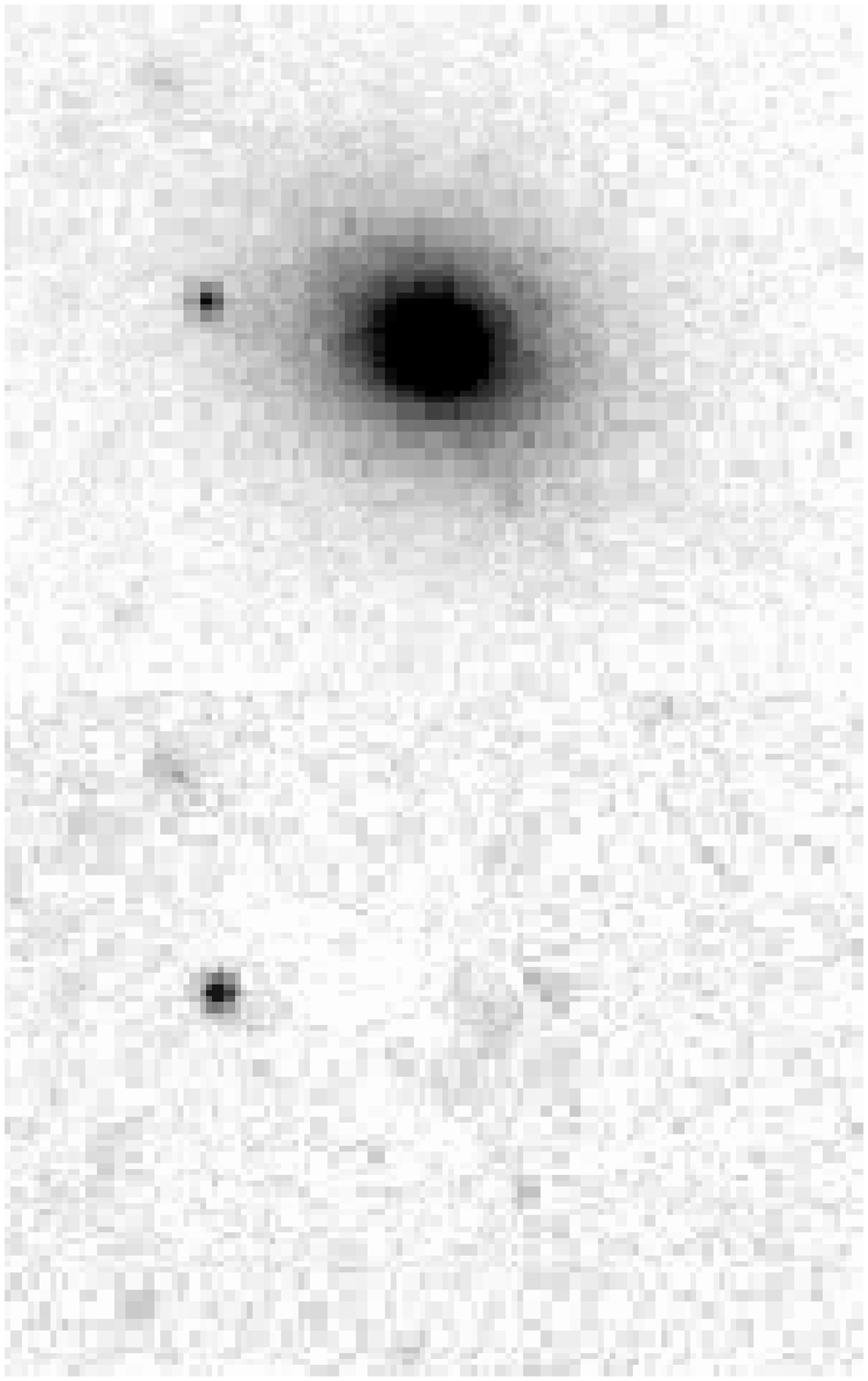}
\plottwo{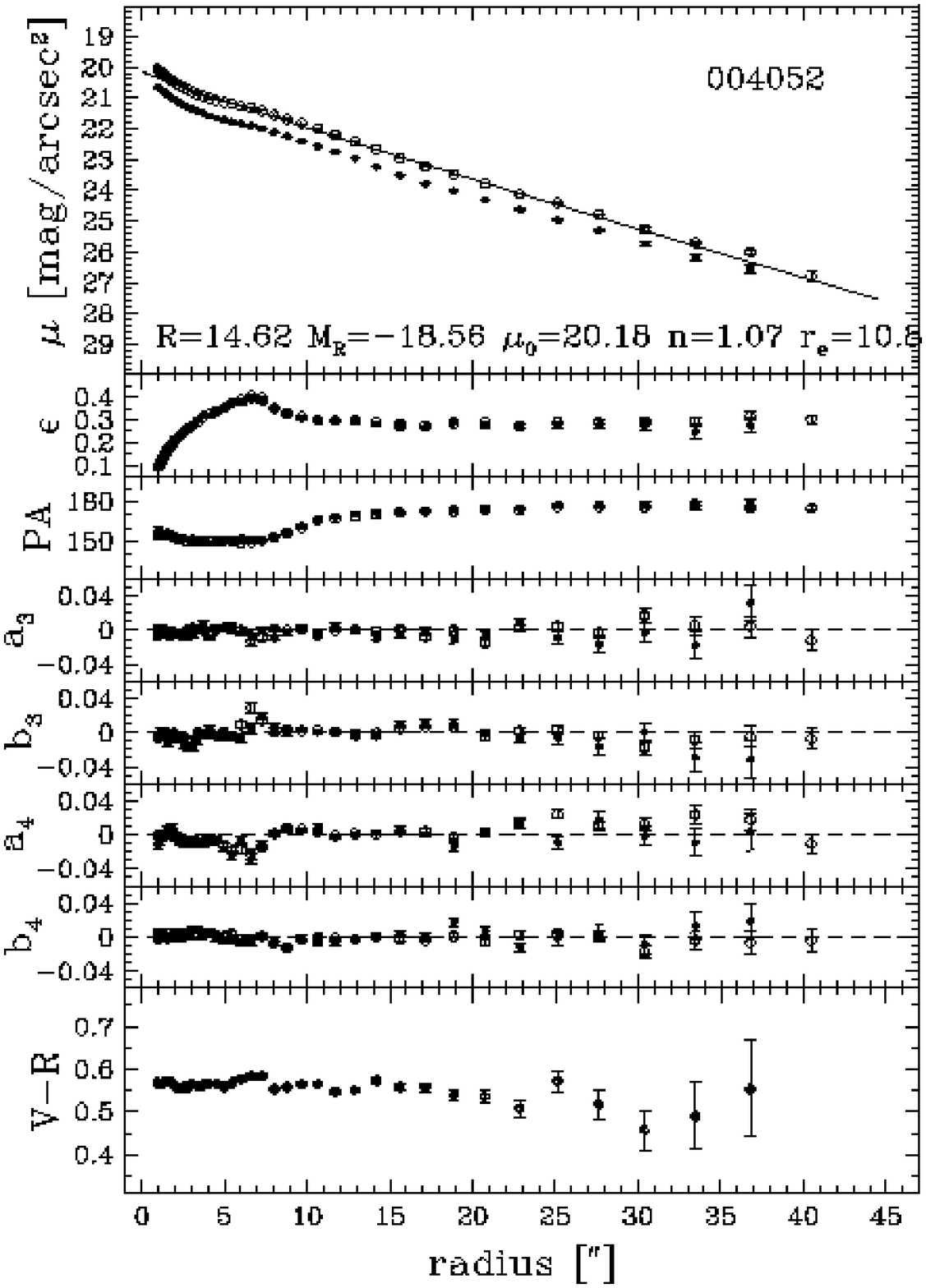}{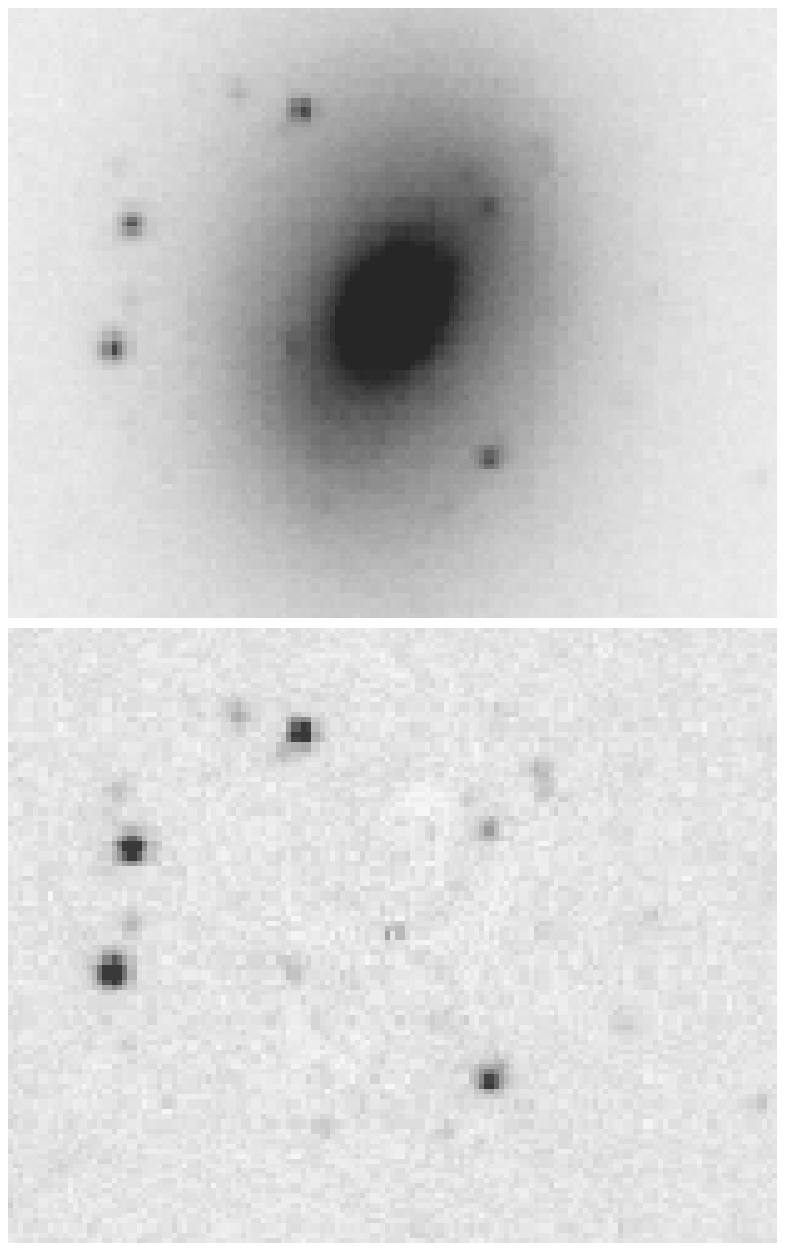}
}
\resizebox{16.7cm}{!}{
\plottwo{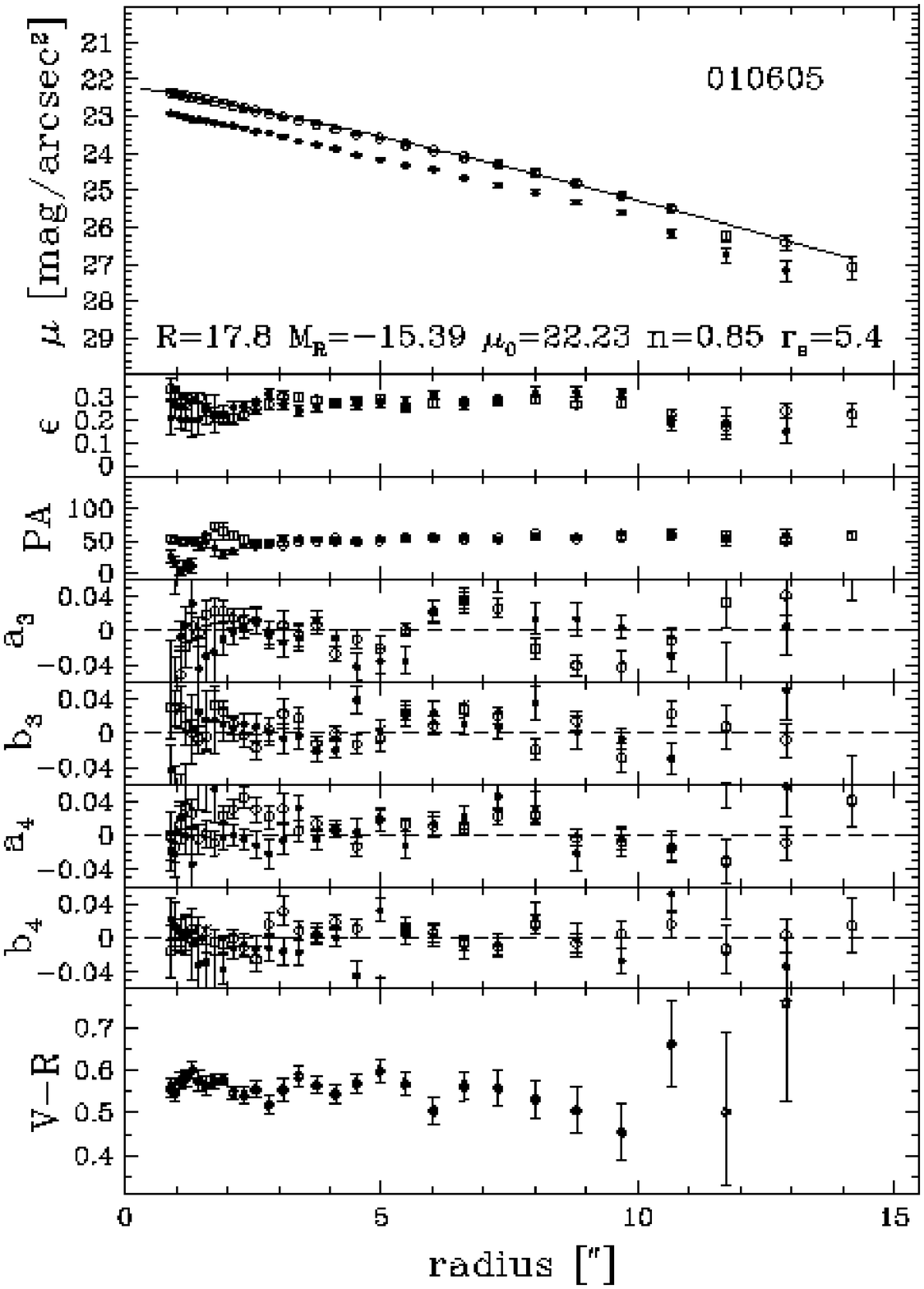}{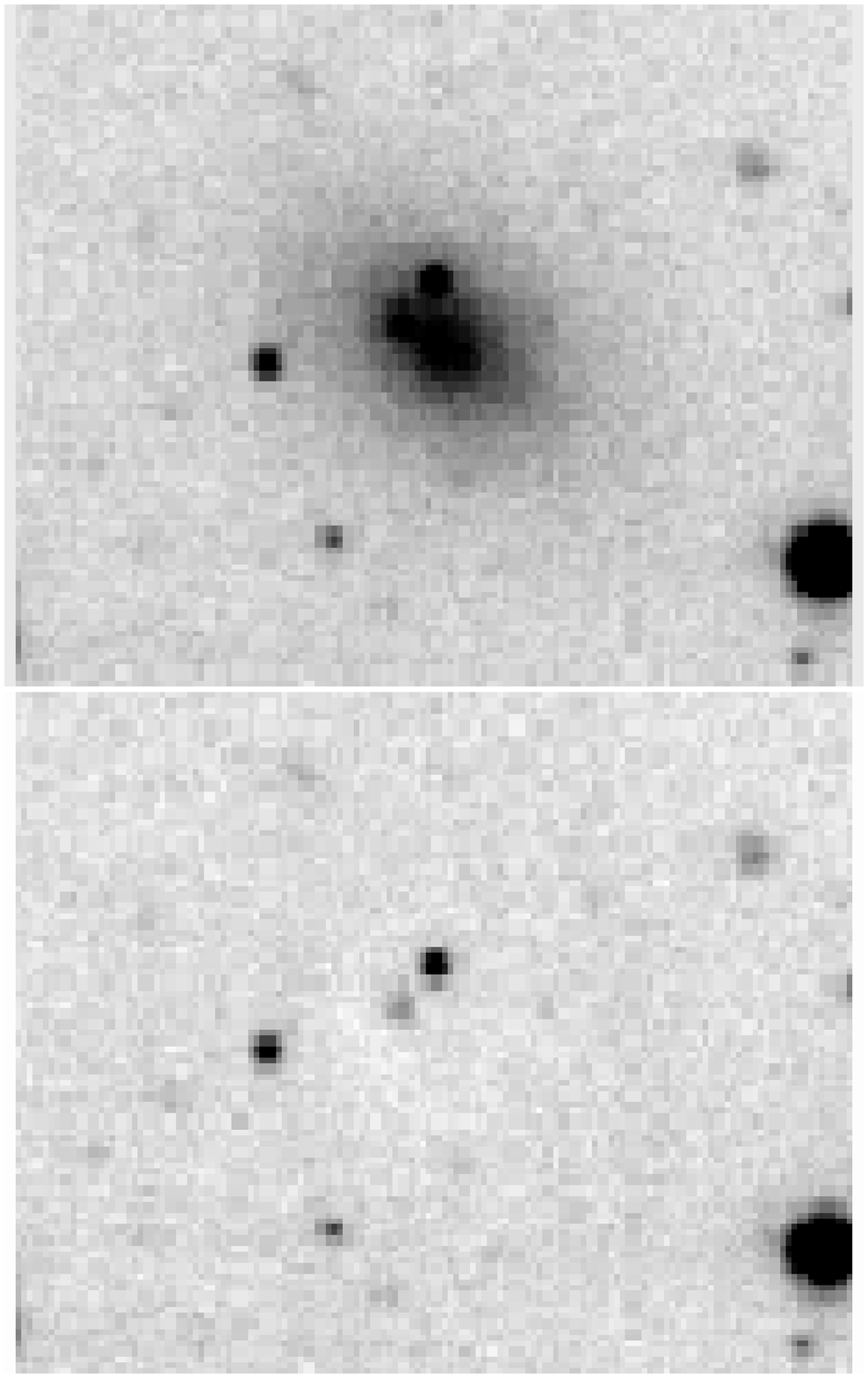}
\plottwo{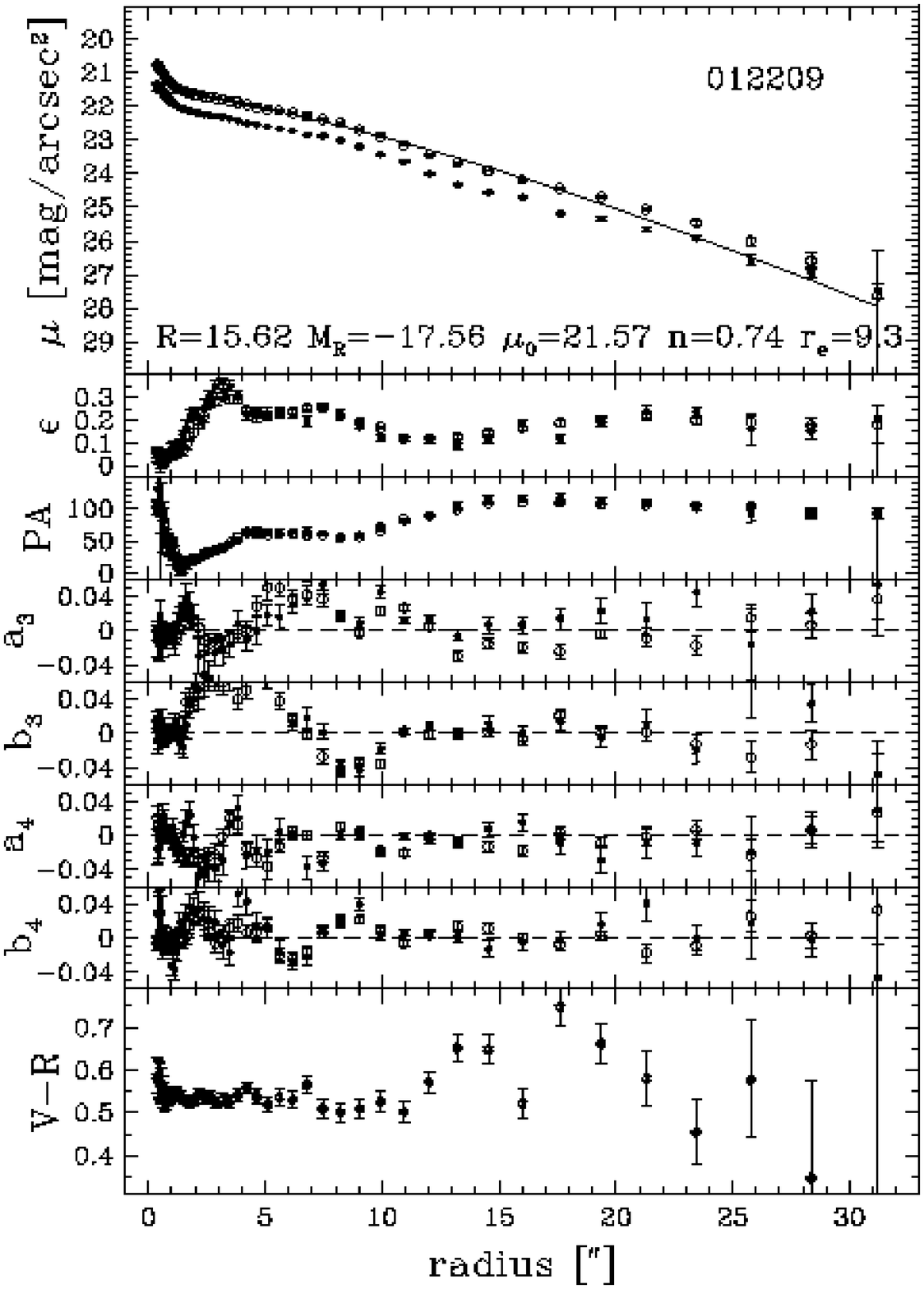}{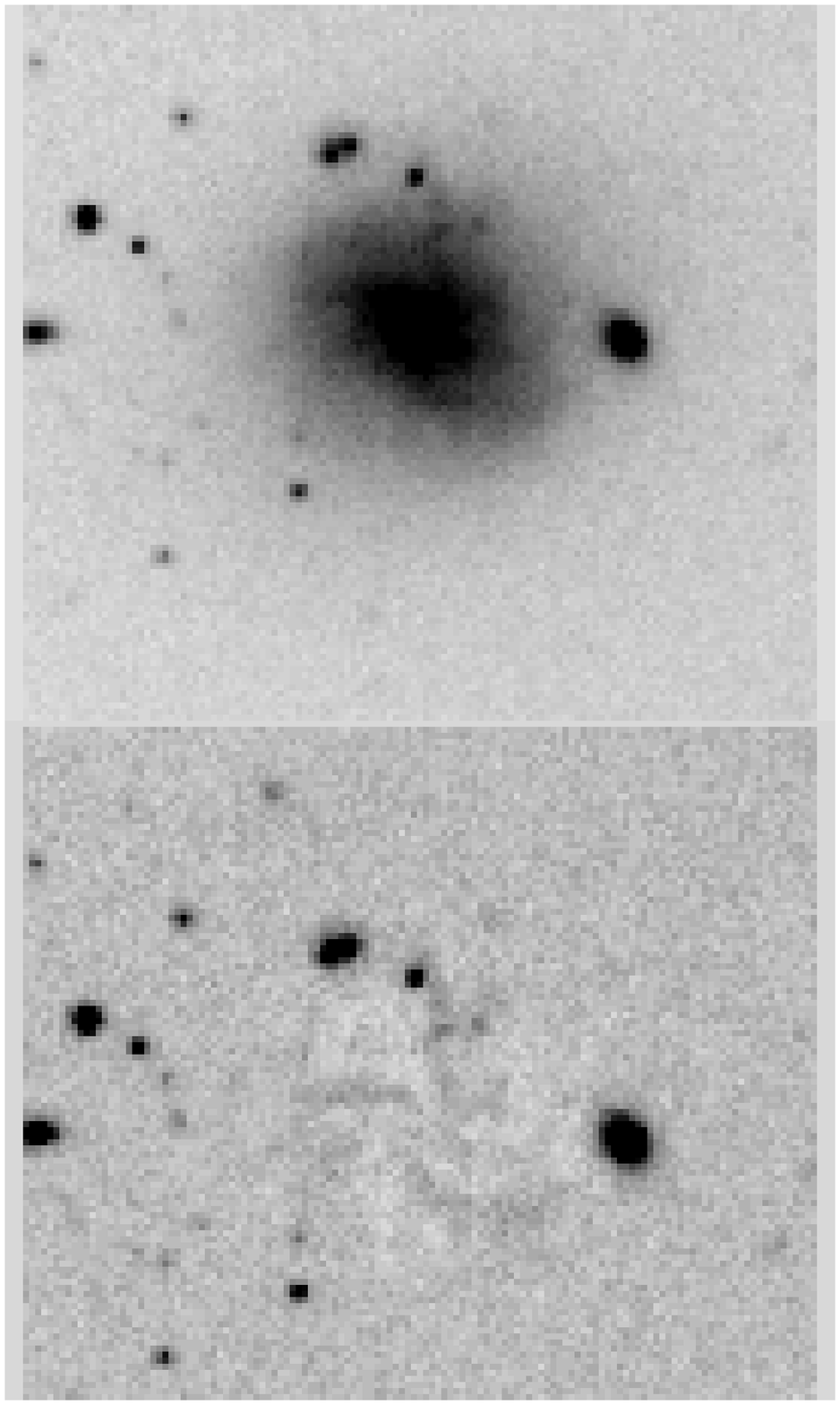}
}
\caption{ A sample of candidate members. Luminosity profiles
obtained with the {\tt IRAF-ELLIPSE} procedure (left panels), R band image and
residual image after automated model subtraction with the {\tt GALFIT} package
(top and bottom right panels).  The model parameters for each object are
provided in the surface brightness panels. The solid line represents the
surface brightness profile of the {\tt GALFIT}-model.
\label{figure14}}
\end{figure*}

\clearpage



\begin{figure}
\epsscale{1}
\plotone{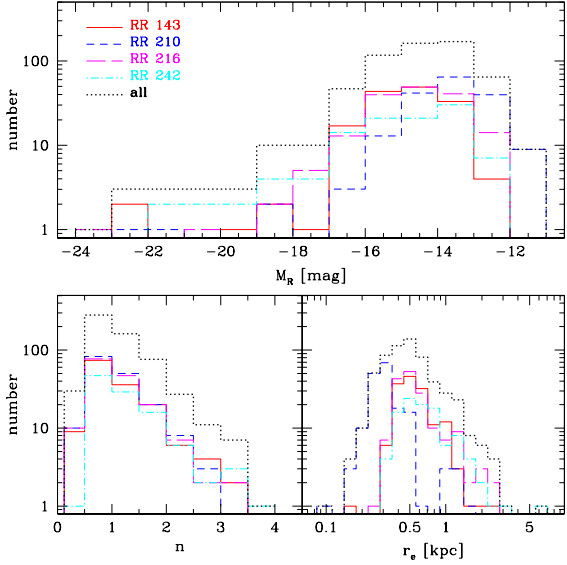}
\caption{  Distribution of absolute R band magnitudes ({\it top}),
effective radii ({\it bottom right}) and Sersic indices n ({\it
bottom left}) of the pair/group members and possibly associated faint
galaxy population.
\label{figure15}}
\end{figure}

\begin{figure}
\epsscale{1}
\caption{  Surface density maps of the candidate faint members selected
according to the color magnitude relation shown in Figure~\ref{figure13}: {\bf
RR~143 (top left), RR~210 (top right), RR~216 (bottom left), RR~242 (bottom
right).} The positions of (bright) confirmed members of each galaxy system are
indicated with a symbol size according to their absolute magnitude (Figures
\ref{figure7} - \ref{figure10} provide the identifications).
\label{figure16}}
\end{figure}

\begin{figure}
\epsscale{1}
\caption{ Relations between central surface brightness $\mu_0$,
Sersic index n and absolute R band magnitude. The bold lines define
the relations for a large sample of early type galaxies taken from
\citet{Gra05}. Large symbols represent spectroscopic confirmed member
galaxies (see Figures \ref{figure7} - \ref{figure10} for identifications).
\label{figure17}}
\end{figure}



\begin{figure*}
\epsscale{1}
\plotone{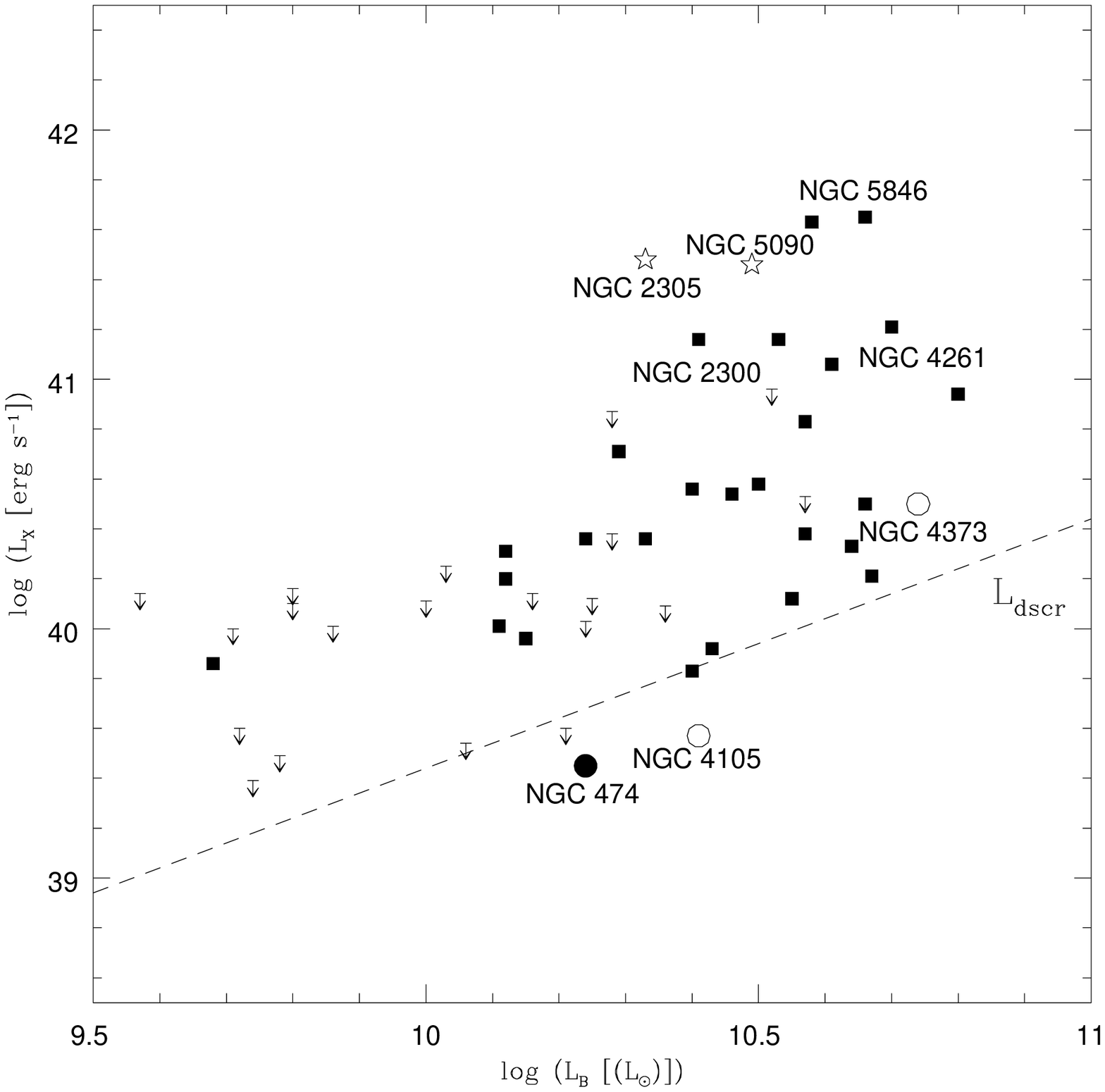}
\plotone{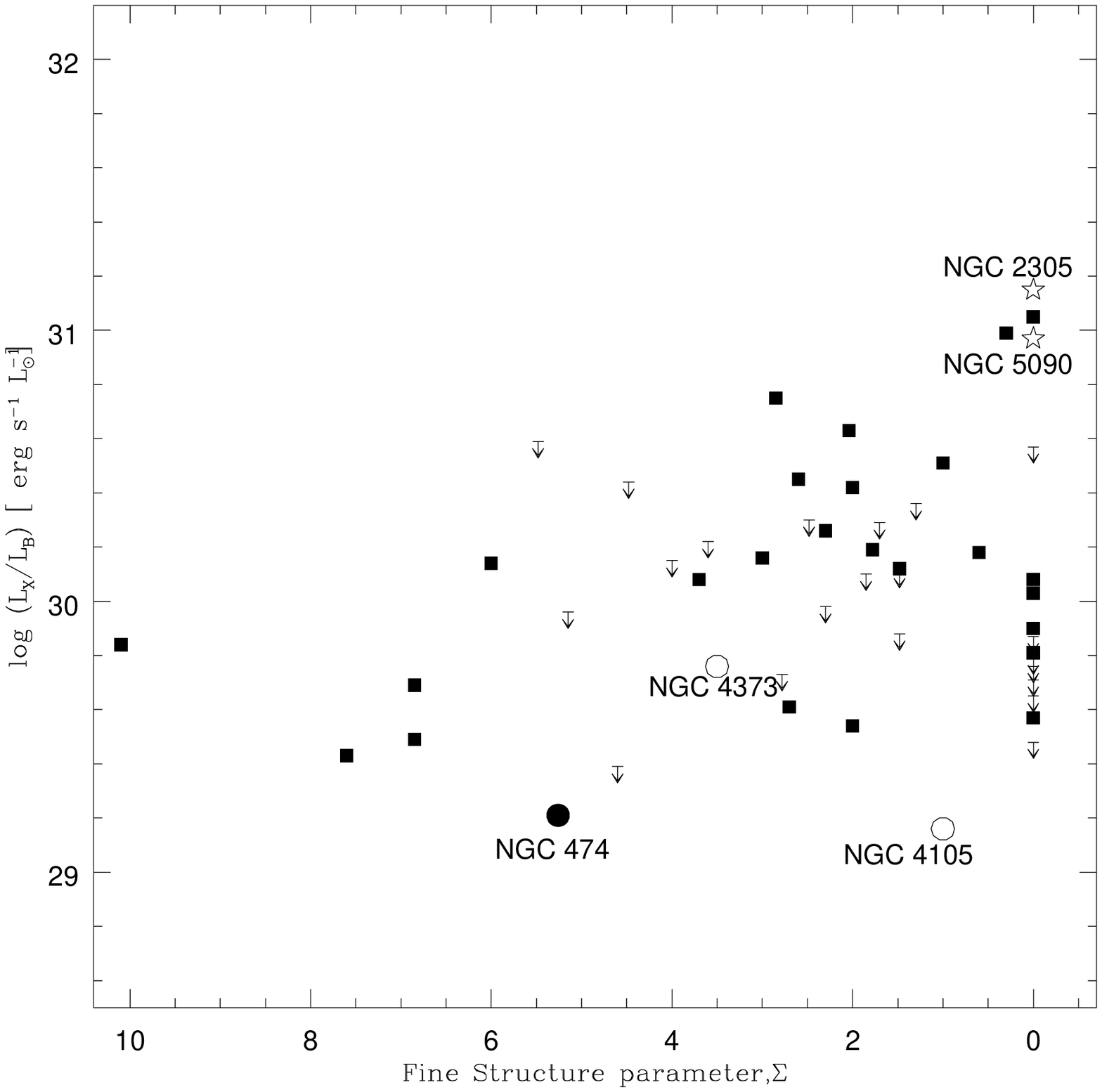}
\caption{(left panel)
Location of the early-type galaxies members of our systems  in the
L$_B$ - L$_X$ plane. With open stars we plot the location of our ETGs observed
with XMM-Newton, with open circles we plot NGC~2305 and NGC~4373 observed with
ROSAT by \citet{TR01}.  For comparison we plot also early-type galaxies with
fine structure, indicative of {\it dynamically young} galaxies, from the
sample of \citet{Sansom00}.  Notice in particular the location of NGC~474
\citep{Rampa06} which has a large system of shells and is interacting with the
spiral NGC~470.  X-ray luminosities are taken from the \citet{Osul01}
compilation. We have also used their recipe to re-calculate the luminosity,
given in Table~\ref{table3}. We have labeled some galaxies in this sample that are members of poor groups \citep[see e.g.][]{Helsdon01}. The dashed line (L$_{dscr}$) represents the expected contribution from discrete stellar X-ray sources from \citet{Ciotti91}. Filled
squares are detections and arrows denote upper limits. (right panel) Position
of galaxies in the plot of normalized X-ray luminosity versus fine structure parameter
$\Sigma$ for early type galaxies.
\label{figure18}}
\end{figure*}


\clearpage


\begin{deluxetable}{rrrrrrrrrc}
\tabletypesize{\scriptsize}
\rotate
\tablecaption{Overview of the pairs and their members}
\tablewidth{0pt}
\tablehead{\colhead{} &  \multicolumn{2}{c}{RR~143} &
\multicolumn{2}{c}{RR~210} &\multicolumn{2}{c}{RR~216}&
\multicolumn{2}{c}{RR~242}\\
\colhead{} & \colhead{NGC~2305} & \colhead{NGC~2307} &
\colhead{NGC~4105} & \colhead{NGC~4106} &
\colhead{IC~3290} & \colhead{NGC~4373} &
\colhead{NGC~5090} & \colhead{NGC~5091} &
\colhead{notes}}
\startdata
ESO ident           &870440   & 870450   & 4400540  & 4400560  &
3220040  & 3220060 & 2700020 & 2700040   &      \\
RA (2000)           &06 48 36  &06 48 50 & 12 06 41  & 12 06 45  & 12
25 09  & 12 25 18 & 13 21 12 & 13 21 18    &       \\
Decl. (2000)        &-64 16 24 & -64 20 01&-29 45 42 & -29 46 06&-39
46 31  &-39 45 37 &-43 42 16&-43 43 10   &   \\
 Morphol. Type   & E2: pec  & SB(rs)ab     & E3  &SB(s)0+ & SB(s)a &
SAB(rs)0-: & E2 & Sb pec sp &  [1]   \\
                         &         &        &      &       &     &
&    &  &\\
\hline
{\bf App. magnitude and color}: &  NGC~2305 & NGC~2307 & NGC~4105 & NGC~4106 & IC~3290 & NGC~4373 & NGC~5090 & NGC~5091 & notes \\
\hline                                              &
&              &  & & & & & &  \\
B$_T$        & 12.74       & 12.92 & 11.60 & 12.36 & 12.96 & 11.90 &
12.57 & 13.91& [1]   \\
\enddata
\tablecomments{In general the photometric, kinematical and structural
data are derived from  {\tt NED} [1],   {\tt LEDA} [2] and  {\tt
Hypercat
http://www-obs.univ-lyon1.fr/hypercat} [3]
compilations. Detailed studies of some objects have been done by our
group and labeled as follows: \citet{RR96} [4], \citet{R88} [5],
\citet{ReRa95} [6], \citet{Long98b} [7], the original ROSAT data published in \citet{TR01} are converted to the 0.5 - 2 keV band and to H$_0$=75 km~s$^{-1}$~Mpc$^{-1}$ [8]. The shape parameter a4/a is indicated as boxy (a4/a$<$0), disky (a4/a $>$0) and irregular (irr)
when there is no definite trend in the shape profile with radius. The adopted
distance is obtained using H$_0$=75 km~s$^{-1}$~Mpc$^{-1}$.
}
\label{table1}
\end{deluxetable}


\clearpage


\begin{deluxetable}{llcc} \tabletypesize{\scriptsize}
\tablecaption{Log-book of XMM-Newton observations. \label{table2}}
\tablewidth{0pt} \tablehead{
\colhead{Pair} & \colhead{Obs. date} & \colhead{Original Time} &
\colhead{Used livetime}\\
\colhead{} & & \colhead{PN/MOS} &\colhead{[s]} \\
}
\startdata
RR~143 &2003-10-19 &9325/11850 & 7289/9250 \\
RR~242 &2003-07-17 &18032/21650 &11794/14318 \\
\enddata
\end{deluxetable}



\begin{deluxetable}{lccccccl} \tabletypesize{\scriptsize}
\tablecaption{Basic X-ray properties of the extended component
\label{table3}}
\tablewidth{0pt} \tablehead{
\colhead{Object } & \multicolumn{2}{c}{extension}  &
\colhead{counts} & \colhead{count } & \colhead{flux} &
\colhead{log L$_X$} & \colhead{Spectral parameters $\dagger$}
\\
\colhead{} & \colhead{[\arcmin]} & \colhead{[kpc]} &
\colhead{[0.5-2.0keV]} & \colhead{rate} & \colhead{[0.5-2.0keV]} &
\colhead{[erg~s$^{-1}$]} & \colhead{kT [90\% confidence]}
\\
}
\startdata
RR~143a (NGC~2305) & 8 & 120 &5220$\pm$95 & 0.72 &1.2$\times 10^{-12}$ &41.48
& 0.54 [0.52-0.56] \\
RR~242a (NGC~5090) & 12 & 160&[5750$\pm$135]$\dagger\dagger$
&0.49 & 1.1$\times 10^{-12}$ &41.46
& 0.3 [0.25-0.37];2.7 [2.4-3.3] \\
\enddata
\tablecomments{\\
$\dagger$ Best fit temperature(s) and 90\% confidence limits assuming a MEKAL code.
Low energy absorption is left free, but it is always consistent with the line of sight
value.  Abundance ratios are consistent with the solar value. \\
$\dagger\dagger$ Estimated excluding the central source.}
\end{deluxetable}


\clearpage


\begin{deluxetable}{lcclcccc} \tabletypesize{\scriptsize}
\tablecaption{Log-book of Wide Field Imaging observations.
\label{table4}}
\tablewidth{0pt} \tablehead{ \colhead{(1)} & \colhead{(2)} & \colhead{(3)}
& \colhead{(4)} & \colhead{(5)} & \colhead{(6)}& \colhead{(7)} & \colhead{(8)} \\ }
\startdata
RR\,143  &  18-11-2001 & V  & N &  $6 \times 500$ & 0.7 & 1.22 & 24.19 \\
RR\,143  &  17-11-2001 & R  & N &  $6 \times 500$ & 0.7 & 1.22 & 24.53 \\

RR\,210   &  17-04-2002 & V  & Y &  $6 \times 500$ & 0.8 & 1.18 & 24.12 \\
RR\,210   &  22-01-2002 & R  & Y &  $6 \times 500$ & 0.7 & 1.03 & 24.48 \\

RR\,216  &  25-01-2002 & V  & Y &  $6 \times 500$ & 0.7 & 1.02 & 24.15 \\
RR\,216  &  17-04-2002 & R  & Y &  $6 \times 500$ & 0.7 & 1.08 & 24.44 \\

RR\,242  &  08-06-2004 & V  & Y &  $6 \times 500$ & 0.9 & 1.13 & 24.06 \\
RR\,242  &  08-06-2004 & R  & Y &  $6 \times 500$ & 0.9 & 1.05 & 24.41 \\
\enddata
\tablenotetext{1}{Pair}
\tablenotetext{2}{Observing nights}
\tablenotetext{3}{Filter}
\tablenotetext{4}{Dither}
\tablenotetext{5}{Exposure time [s]}
\tablenotetext{6}{Seeing FWHM [arcsec]}
\tablenotetext{7}{Airmass}
\tablenotetext{8}{Zero Point}
\end{deluxetable}



\clearpage
\begin{deluxetable}{lllcccccc} 
\tabletypesize{\scriptsize}
\tablecaption{Faint candidate group member population around RR~143.
\label{table5}}
\tablewidth{0pt} 
\tablehead{ 
\colhead{ID} & \colhead{$\alpha$} & \colhead{$\delta$} & \colhead{m$_R$} &
\colhead{V--R} & \colhead{n} & \colhead{r$_e$} & \colhead{$\mu_0$} & \colhead{$\mu_{e}$}
\\
\colhead{} & \colhead{} & \colhead{} & \colhead{[mag]} &  \colhead{[mag]} & \colhead{} & \colhead{[arcsec]} &  \colhead{[mag/arcsec$^2$]} & \colhead{[mag/arcsec$^2$]} \\
}
\startdata

00006  &  06  47  20.3  &  -64  31  19  &  15.23  &  0.56  &  3.17  &  6.77  &  15.66  &  22.18  \\
00028  &  06  49  06.8  &  -64  31  42  &  19.27  &  0.84  &  1.08  &  1.63  &  20.35  &  22.35  \\
00035  &  06  47  51.2  &  -64  31  48  &  18.04  &  0.62  &  1.53  &  1.58  &  18.89  &  21.86  \\
00751  &  06  47  49.6  &  -64  00  22  &  19.51  &  0.46  &  0.67  &  1.94  &  21.23  &  22.33  \\
00754  &  06  46  07.3  &  -64  00  19  &  18.66  &  0.59  &  0.93  &  1.81  &  20.10  &  21.76  \\
01042  &  06  47  10.1  &  -64  02  21  &  14.36  &  0.69  &  1.42  &  8.11  &  17.90  &  20.63  \\
01127  &  06  50  53.6  &  -64  00  45  &  19.97  &  0.25  &  0.49  &  2.22  &  22.71  &  23.43  \\
01326  &  06  50  59.0  &  -64  01  02  &  18.81  &  0.52  &  0.52  &  2.71  &  21.50  &  22.26  \\
\enddata
\end{deluxetable}


\clearpage



\clearpage

\begin{deluxetable}{lllcccccc} 
\tabletypesize{\scriptsize}
\tablecaption{Faint candidate group member population around RR~210.
\label{table6}}
\tablewidth{0pt} 
\tablehead{ 
\colhead{ID} & \colhead{$\alpha$} & \colhead{$\delta$} & \colhead{m$_R$} &
\colhead{V--R} & \colhead{n} & \colhead{r$_e$} &  \colhead{$\mu_0$} & \colhead{$\mu_{e}$}
\\
\colhead{} & \colhead{} & \colhead{} & \colhead{[mag]} &  \colhead{[mag]} & \colhead{} & \colhead{[arcsec]} & \colhead{[mag/arcsec$^2$]} & \colhead{[mag/arcsec$^2$]} \\
}
\startdata

00129  &  12  07  09.0  &  -30  00  26  &  18.94  &  0.55  &  0.78  &  2.03  &  21.32  &  22.67  \\
00232  &  12  05  47.2  &  -30  00  15  &  18.59  &  0.66  &  0.97  &  1.47  &  20.49  &  22.24  \\
00371  &  12  05  52.4  &  -29  59  58  &  20.10  &  0.59  &  0.50  &  2.27  &  22.46  &  23.20  \\
00554  &  12  06  46.1  &  -29  59  49  &  18.06  &  0.50  &  0.73  &  1.89  &  20.51  &  21.75  \\
00568  &  12  06  43.5  &  -30  00  00  &  18.03  &  0.58  &  1.79  &  2.51  &  19.41  &  22.95  \\
00654  &  12  05  50.2  &  -29  59  58  &  17.39  &  0.66  &  1.84  &  2.41  &  18.30  &  21.93  \\
00889  &  12  06  58.9  &  -29  59  45  &  17.55  &  0.72  &  1.35  &  1.64  &  18.99  &  21.56  \\
00910  &  12  07  41.4  &  -29  59  21  &  18.57  &  0.49  &  0.73  &  2.12  &  20.73  &  21.96  \\
\enddata
\end{deluxetable}


\clearpage



\clearpage

\begin{deluxetable}{lllcccccc} 
\tabletypesize{\scriptsize}
\tablecaption{Faint candidate group member population around RR~216.
\label{table7}}
\tablewidth{0pt} 
\tablehead{
\colhead{ID} & \colhead{$\alpha$} & \colhead{$\delta$} & \colhead{m$_R$} & 
\colhead{V--R} & \colhead{n} & \colhead{r$_e$} & \colhead{$\mu_0$} & \colhead{$\mu_{e}$}
\\
\colhead{} & \colhead{} & \colhead{} & \colhead{[mag]} & \colhead{[mag]} &  \colhead{} & \colhead{[arcsec]} & \colhead{[mag/arcsec$^2$]} & \colhead{[mag/arcsec$^2$]} \\
}
\startdata

00015  &  12  24  10.1  &  -39  59  31  &  18.47  &  0.47  &  0.74  &  1.87  &  20.43  &  21.70  \\
00017  &  12  24  30.0  &  -39  59  28  &  16.20  &  0.59  &  1.03  &  6.54  &  20.37  &  22.25  \\
00034  &  12  25  12.4  &  -39  59  50  &  19.29  &  0.49  &  0.65  &  1.88  &  21.49  &  22.56  \\
00143  &  12  24  29.0  &  -39  29  21  &  17.19  &  1.03  &  1.17  &  1.64  &  18.83  &  21.02  \\
00193  &  12  24  31.8  &  -39  29  33  &  18.06  &  0.72  &  1.72  &  1.74  &  18.59  &  21.98  \\
00226  &  12  25  20.0  &  -39  29  36  &  20.03  &  0.60  &  1.27  &  5.74  &  23.21  &  25.62  \\
00293  &  12  23  54.8  &  -39  29  41  &  18.75  &  0.76  &  0.71  &  3.03  &  21.44  &  22.63  \\
00495  &  12  25  46.1  &  -39  29  54  &  18.81  &  0.28  &  1.21  &  1.95  &  20.20  &  22.47  \\
\enddata
\end{deluxetable}


\clearpage



\clearpage

\begin{deluxetable}{lllcccccc} 
\tabletypesize{\scriptsize}
\tablecaption{Faint candidate group member population around RR~242.
\label{table8}}
\tablewidth{0pt} 
\tablehead{ 
\colhead{ID} & \colhead{$\alpha$} & \colhead{$\delta$} & \colhead{m$_R$} &
\colhead{V--R} & \colhead{n} & \colhead{r$_e$} & \colhead{$\mu_0$} & \colhead{$\mu_{e}$}
\\
\colhead{} & \colhead{} & \colhead{} & \colhead{[mag]} & \colhead{[mag]} & \colhead{} & \colhead{[arcsec]} & \colhead{[mag/arcsec$^2$]} & \colhead{[mag/arcsec$^2$]} \\
}
\startdata

00017  &  13  20  19.9  &  -43  55  57  &  17.03  &  0.69  &  3.05  &  3.58  &  16.44  &  22.72  \\
00018  &  13  21  39.0  &  -43  55  54  &  17.85  &  0.77  &  2.46  &  1.90  &  16.98  &  21.97  \\
00636  &  13  22  22.3  &  -43  27  10  &  18.63  &  0.43  &  1.30  &  2.30  &  20.89  &  23.36  \\
00842  &  13  22  33.7  &  -43  26  06  &  20.32  &  0.49  &  0.61  &  1.99  &  22.43  &  23.41  \\
00970  &  13  19  47.3  &  -43  27  32  &  19.54  &  0.64  &  0.71  &  1.84  &  21.59  &  22.79  \\
01414  &  13  21  38.1  &  -43  26  20  &  16.50  &  0.63  &  1.96  &  1.68  &  16.45  &  20.35  \\
02287  &  13  20  21.5  &  -43  28  23  &  20.46  &  0.47  &  0.60  &  1.89  &  21.98  &  22.93  \\
02474  &  13  20  19.8  &  -43  27  48  &  16.66  &  0.62  &  2.99  &  2.67  &  15.81  &  21.95  \\
\enddata
\end{deluxetable}


\clearpage



\clearpage
   \begin{deluxetable}{lrrrrrl}
\tabletypesize{\scriptsize}
\tablecaption{X-ray sources in the RR~143 field}
\tablewidth{0pt}
\tablehead{
\colhead{N }   &  \colhead{RA}      &   \colhead{DEC }   &  \colhead{Net$^\dagger$ } &\colhead{Cnt$^\dagger$ }
 &\colhead{flux$^\dagger$} & \colhead{Instr}\\
 &&& \colhead{counts} &\colhead{rate} &\colhead{(erg s$^{-1}$)} }
\startdata
    ~1  & 06 46 40.8 & -64 16 02 & 26.71 $\pm$6.18 & 0.0104 & 1.013168E-13 & PN \\ 
    ~2  & 06 47 16.0 & -64 23 17 & 47.95 $\pm$8.65 & 0.0519 & 7.104581E-13 & PN \\ 
    ~3  & 06 47 24.0 & -64 23 32 & 78.73 $\pm$10.07 & 0.0256 & 2.441722E-13 & PN \\ 
    ~4  & 06 47 27.4 & -64 04 56 & 86.03 $\pm$9.97 & 0.0355 & 3.905863E-13 & PN \\ 
    ~5  & 06 47 34.4 & -64 18 34 & 22.64 $\pm$6.15 & 0.0053 & 5.150544E-14 & PN \\ 
    ~6  & 06 47 40.0 & -64 06 53 & 23.01 $\pm$6.16 & 0.0073 & 7.321936E-14 & PN \\ 
    ~7  & 06 47 48.7 & -64 19 56 & 25.63 $\pm$5.95 & 0.0055 & 6.401829E-14 & PN \\ 
    ~8  & 06 47 55.9 & -64 07 35 & 17.68 $\pm$4.68 & 0.0056 & 4.716441E-14 & PN \\
\enddata
  \tablecomments{$^\dagger$ In the full 0.5-10.0 keV band. Fluxes are derived for a power law spectrum
  with $\Gamma=1.7$ and galactic absorption. \\
  $^{\dagger\dagger}$ RR~143a \\
  $^{\dagger\dagger\dagger}$ Companion RR~143b [NGC~2307]  }
 \label{table143}
 \end{deluxetable}

\clearpage



\clearpage

     \begin{deluxetable}{lrrrrrl}
\tabletypesize{\scriptsize}
\tablecaption{X-ray sources in the RR~242 field}
\tablewidth{0pt}
\tablehead{
\colhead{N }   &  \colhead{RA}      &   \colhead{DEC }   &  \colhead{Net$^\dagger$ } &\colhead{Cnt$^\dagger$ }
 &\colhead{flux$^\dagger$} & \colhead{Instr}\\
 &&& \colhead{counts} &\colhead{rate} &\colhead{(erg s$^{-1}$)} }
\startdata
    ~1  & 13 20 00.0 & -43 46 40 & 50.84 $\pm$8.02 & 0.0102 & 3.082375E-13 & MOS \\ 
     ~2  & 13 20 12.2 & -43 40 12 & 45.10 $\pm$8.82 & 0.0082 & 1.227895E-13 & PN \\ 
     ~3  & 13 20 18.7 & -43 33 56 & 79.76 $\pm$10.21 & 0.0130 & 3.278396E-13 & MOS \\ 
     ~4  & 13 20 20.5 & -43 51 24 & 38.20 $\pm$8.09 & 0.0094 & 1.507757E-13 & PN \\ 
     ~5  & 13 20 22.6 & -43 32 59 & 33.81 $\pm$6.99 & 0.0056 & 2.065168E-13 & MOS \\ 
     ~6  & 13 20 29.8 & -43 34 51 & 39.48 $\pm$7.68 & 0.0066 & 7.311842E-14 & PN \\ 
     ~7  & 13 20 30.3 & -43 41 44 & 30.38 $\pm$7.96 & 0.0040 & 6.720174E-14 & PN \\ 
     ~8  & 13 20 40.3 & -43 42 00 & 305.49 $\pm$20.70 & 0.0323 & 3.203940E-13 & PN \\
\enddata
  \tablecomments{$^\dagger$ In the full 0.5-10.0 keV band. Fluxes are derived for a power law spectrum
  with $\Gamma=1.7$ and galactic absorption.
  }
 \label{table242}
 \end{deluxetable}

\clearpage



\begin{thebibliography}{}

\bibitem[Annibali et al.(2006)]{Annibali06} Annibali, F., Bressan, A., Rampazzo, R.,
Danese, L. \& Zeilinger, W. W. 2006, A\&A, submitted.

\bibitem[Arp \& Madore(1987)]{AM87} Arp, H. C. \& Madore, B.F. 1987,
Catalogue
of Southern Peculiar Galaxies and Associations, V. 1-2, Cambridge
University Press

\bibitem[Barnes (1996)]{Barnes96} Barnes, J. 1996, Galaxies:
Interactions an Induced Star Formation (26th Saas-Fee Advanced Course)
(New York: Springer), 275

\bibitem[Belsole et al.(2003)]{Bel03} Belsole, E., Sauvageot, J.-L.,
Ponman, T. J. \& Bourdin, H.  2003, AA, 398, 1

\bibitem[Bertin et al.(1996)]{Bert96} Bertin, E. \& Arnouts, S. 1996, AAS, 117, 393

\bibitem[Bettoni et al.(2003)]{Bettoni03} Bettoni, D., Falomo, R.,
Fasano, G. \& Govoni, F. 2003, AA, 399, 869

\bibitem[Beuing et al.(2002)]{Beuing} Beuing, J., Bender, R.,
Mendes de Oliveira, C., Thomas, D. \& Maraston, C. 2002 A\&A, 395, 431

\bibitem[Binggeli, Sandage \& Tarenghi(1984)]{Bin84} Binggeli, B., Sandage,
  A.\& Tarenghi, M. 1984, AJ, 89, 64

\bibitem[Caon et al.(2000)]{Caon00} Caon, N., Macchetto, D. \&
Pastoriza, M. 2000, ApJS, 127, 39

\bibitem[Cappellari et al. (2006)]{Cappellari05} Cappellari, M., Bacon, R.,
Bureau M., Damen, M. C., Davies, R. L., de Zeeuw, P. T., Emsellem, E.,
Falcon-Barroso, J., Krajnovic, D., Kuntschner, H., McDermid, R. M.,
Peletier, R. F., Sarzi M., van den Bosch, R. C. E. \& van de Ven, G. 2006,
MNRAS, 366, 1151

\bibitem[Carollo et al.(1993)]{Carol93} Carollo, M., Danziger, I. J.
\& Buson, L. M. 1993, MNRAS, 265, 553

\bibitem[Chapman et al. (2004)]{Chapman04} Chapman, S.C., Smail, I.,
Blain, A. W. \& Ivison, R. J. 2004, ApJ, 614, 671

\bibitem[Ciotti et al.(1991)]{Ciotti91} Ciotti, L., Pellegrini, S.,
Renzini, A. \& D'Ercole, A. 1991, ApJ, 376, 380

\bibitem[Clemens et al.(2006)]{Clemens06} Clemens, M. S., Bressan, A.,
Nikolic, B., Alexander, A., Annibali, F. \& Rampazzo, R. 2006, MNRAS,
submitted

\bibitem[Colbert et al.(2001)]{Colbert01} Colbert, J. W., Mulchaey, J. S.
\& Zabludoff, A. I. 2001, AJ, 121, 808

\bibitem[De Lucia et al.(2006)]{DeLucia06} De Lucia G., Springel, V.,
White, S. D. M., Croton, D. \& Kauffmann, G. 2006, MNRAS, 366, 499

\bibitem[De Souza et al.(2004)]{DeSou04} De Souza, R.E., Gadotti,
D.A. \& Dos Anjos S. 2004, ApJS, 153, 411

\bibitem[Dickens et al.(1986)]{Dick86} Dickens, R. J., Currie, M. J.
\& Lucey, J. R. 1986, MNRAS, 220, 679

\bibitem[Domingue et al.(2005)]{Dom05} Domingue, D. L., Sulentic, J. W.
\& Durbala, A. 2005, AJ, 129, 2579

\bibitem[Domingue et al.(2003)]{Dom2003} Domingue, D. L., Sulentic, J. W.,
Xu, C., Mazzarella, J., Gao Y. \& Rampazzo, R. 2003, AJ, 125, 555

\bibitem[Dressler(1980)]{Dre80} Dressler, A. 1980, ApJ, 236, 351

\bibitem[Dressler(1984)]{Dre84} Dressler, A. 1984, ApJ, 281, 512

\bibitem[Ellis \& Bland-Hawthorn(2006)]{Ellis06} Ellis, S.C. \& Bland-Hawthorn, J. 2006,
astro-ph/06025573

\bibitem[Eracleous \& Halpern(2001)]{Eracl01} Eracleous, M. \& Halpern, J. P.
2001, ApJ, 554, 240

\bibitem[Fabbiano \& Schweizer(1995)]{Fabbiano95} Fabbiano, G., \& Schweizer,
  F. 1995, ApJ, 447, 572

\bibitem[Faber(1973)]{Fab73} Faber, S. M. 1973, ApJ, 179, 731

\bibitem[Faber et al.(1989)]{Faberetal} Faber, S. M., Wegner, G.
Burstein, D., Davies, R. L., Dressler, A., Lynden-Bell, D. \&
Terlevich, R. J. 1989, ApJS, 69, 763

\bibitem[Falcon-Barroso et al.(2005)]{Falcon05} Falcon-Barroso, J. et al. 2005,
NewAr, 49, 515

\bibitem[Franx et al.(2003)]{Franx03} Franx, M. et al. 2003, ApJ, 587, L79

\bibitem[Gonzales et. al(2005)]{Gon05} Gonzales, A.H., Zabludoff, A. I.
\&  Zaritsky, D. 2005, ApJ, 618, 195

\bibitem[Govoni et al.(2000)]{Govo00} Govoni, F., Falomo, R., Fasano,
G. \& Scarpa, R. 2000, AAS, 143, 369

\bibitem[Graham(2005)]{Gra05} Graham, A. W. 2005, in Near-fields cosmology with
dwarf elliptical galaxies, IAU Colloquium No. 198, H. Jerjen and B.
Binggeli eds., Cambridge University Press 2005, pp.303-310

\bibitem[Gr\" utzbauch et al.(2005a)]{Gru05a} Gr\" utzbauch, R., Kelm, B., Focardi, P.,
Trinchieri, G., Rampazzo, R. \& Zeilinger, W. W. 2005, AJ, 129, 1832.

\bibitem[Gr\" utzbauch et al.(2005b)]{Gru05b} Gr\" utzbauch, R., Annibali, F., Bressan, A.,
Focardi, P., Kelm, B., Rampazzo, R. \& Zeilinger, W. W. 2005, MNRAS, 364, 146.

\bibitem[Hamabe \& Kormendy(1987)]{HK87} Hamabe, M. \& Kormendy, J. 1987,
IAUS, 127, 379

\bibitem[Helsdon et al.(2001)]{Helsdon01} Helsdon, S.F., Ponman, T.J.,
O'Sullivan, E. \& Forbes, D.A. 2001, MNRAS, 325, 693

\bibitem[Henriksen \& Cousineau(1999)]{HC99} Henriksen, M. \& Cousineau,
S. 1999, ApJ, 511, 595

\bibitem[Hernandez-Toledo et al.(2001)]{HT01}  Hernandez--Toledo, H.
M., Dultzin-Hacyan, D. \&  Sulentic, J. W. 2001, AJ, 121, 1319

\bibitem[Hibbard et al.(1994)]{Hibbard94} Hibbard, J. E., Guhathakurta, P.,
van Gorkom, J. H. \& Schweizer, F. 1994, AJ, 107, 67

\bibitem[Hibbard \& van Gorkom(1996)]{Hibbard96} Hibbard, J. E. \& van Gorkom,
  J. 1996, AJ, 111, 655

\bibitem[Ho et al.(1997)]{Ho97} Ho, L.C., Filippenko, A. V., Sargent W.L.W.
\&  Peng, C.Y. 1997, ApJS, 112, 391

\bibitem[Ho \& Ulvestad(2001)]{Ho01} Ho, L.C. \& Ulvestad, J.S. 2001, ApJS, 133, 77

\bibitem[Jedrzejewski(1987)]{Jedr87} Jedrzejewski, R. 1987, MNRAS, 226, 747

\bibitem[Jones et al.(2003)]{Jon03} Jones et al. 2003, MNRAS, 343, 627

\bibitem[Karachentsev (1987)]{Kara87} Karachentsev, I.D. 1987, Dvoinye
Galaktiki, (Nauka Moskow)

\bibitem [Khosroshahi et al.(2004)]{Kho04} Khosroshahi, H. G.
Raychaudhury,  S., Ponman, T. J., Miles, T. A. \&  Forbes, D. A., 2004, MNRAS, 349, 527

\bibitem[Landolt (1992)]{land92} Landolt, A. U. 1992, \aj, 104, 340

\bibitem[Lloyd et al. (1996)]{Lloyd96} Lloyd, B.D., Jones, P.A.
\& Haynes, R.F. 1996, MNRAS, 279, 1187

\bibitem[Longhetti et al.(1998a)]{Long98a} Longhetti, M., Rampazzo, R.,
Bressan, A. \& Chiosi C.  1998, AAS, 130, 251

\bibitem[Longhetti et al.(1998b)]{Long98b} Longhetti, M., Rampazzo, R.,
Bressan, A. \& Chiosi C.  1998, AAS, 130, 267

\bibitem[Longhetti et al.(1999)]{Long99} Longhetti, M., Bressan, A.,
Chiosi, C. \& Rampazzo, R. 1999, AA, 345, 419

\bibitem[Llyod et al.(1996)]{Llyod96} Llyod, B.D., Jones, P.A. \& Haynes, R.F. 1996, MNRAS,
279, 1187

\bibitem[Meza et al.(2003)]{mez03} Meza, A., Navarro, J. F.,
Steinmetz, M. \& Eke, V. R, 2003, ApJ, 590, 619


\bibitem[Miller et al.(2002)]{Miller02} Miller, N. A., Ledlow, M. J.,
Owen, F. N. \& Hill, J. M. 2002 AJ, 123, 3018

\bibitem[Mulchaey et al.(1993)]{M93} Mulchaey, J. S., Davis, D. S.,
Mushotzky, R.F. \& Burstein, D. 1993, ApJ, 404, L9

\bibitem[Mulchaey et al.(1996)]{M96} Mulchaey, J. S., Davis, D.S.,
Mushotzky, R.F. \& Burstein, D.  1996, ApJ, 456, 80

\bibitem[Mulchaey \& Zabludoff(1999)]{Mul99} Mulchaey, J. S. \&
Zabludoff, A.I. 1999, ApJ, 514, 133

\bibitem[Mulchaey (2000)]{M00} Mulchaey, J. S. 2000, ARAA, 38, 289

\bibitem[Mulchaey et al. (2003)]{Mul03} Mulchaey, J. S., Davis, D.
S., Mushotzky, R. F. \& Burstein, D., 2003, ApJS, 145, 39

\bibitem[Nevalainen et al. (2005)]{Nevalien0X} Nevalainen, J.,
Markevitch, M. \& Lumb, D. 2005, ApJ, 629, 172

\bibitem[Nipoti et al.(2003)]{nip03} Nipoti, C., Londrillo, P.
\& Ciotti, L. 2003, MNRAS, 342, 501

\bibitem[Noguchi (1987)]{No87} Noguchi, M. 1987, MNRAS, 228, 635

\bibitem[Noguchi (1988)]{Nogu88} Noguchi, M. 1988, AA, 203, 259

\bibitem[Nolan et al.(2004)]{Nol04} Nolan, L. A., Ponman, T. J., Read,
A. M. \& Schweizer, F. 2004, MNRAS, 353, 221.

\bibitem[O'Sullivan et al.(2001)]{Osul01} O'Sullivan, E., Forbes, D.A.
\& Ponman, T. J. 2001, MNRAS, 328, 461

\bibitem[Ota et al.(2004)]{Ota04} Ota, N., Morita, U.,
Kitayama, T. \& Ohashi, T.  2004, PASJ, 56, 753

\bibitem[Peng et al.(2002)]{Peng02} Peng, C. Y., Ho, L. C., Impey, C.
D. \& Rix, H. 2002, AJ, 124, 266

\bibitem[Rampazzo (1988)]{R88} Rampazzo, R. 1988, AA, 204, 81

\bibitem[Rampazzo \& Sulentic(1992)]{RS92} Rampazzo, R. \& Sulentic, J. W.
1992, AA, 259, 43

\bibitem[Rampazzo et al.(1998)]{RR98} Rampazzo, R., Covino, S.,
Trinchieri, G. \& Reduzzi, L. 1998, AA, 330, 423

\bibitem[Rampazzo et al.(2006)]{Rampa06} Rampazzo, R., Alexander, P.,
Carignan, C. et al. 2006, MNRAS, in press

\bibitem[Read \& Ponman(1998)]{Read98} Read, A.M. \& Ponman, T.J. 1998,
MNRAS, 297, 143

\bibitem[Read \& Ponman(2003)]{Read0X} Read, A.M. \& Ponman, T.J. 2003,
AA, 409, 395

\bibitem[Reduzzi \& Rampazzo(1995)]{ReRa95} Reduzzi, L. \& Rampazzo, R.
1995, ApLC, 30, 1

\bibitem[Reduzzi \& Rampazzo(1996)]{RR96} Reduzzi, L.\& Rampazzo, R.
1996, AAS, 116, 515

\bibitem[Reduzzi et al.(1996)]{Reduzzi96} Reduzzi, L., Longhetti, M. \& Rampazzo, R.
1996, MNRAS, 282, 149

\bibitem[Rizzi (2003)]{rizzi-phd} Rizzi, L. 2003, Research Doctorate
Thesis, University of Padua

\bibitem[Rose (1984)]{Rose84} Rose, J.A. 1984, AJ, 89, 1238

\bibitem[Rose (1985)]{Rose85} Rose, J.A. 1985, AJ, 90, 1927

\bibitem[Sadler \& Sharp(1984)]{Sadler84} Sadler, E. M. \& Sharp, N. A.
1984, AA, 133, 216

\bibitem[Salo (1991)]{Salo91} Salo, H. 1991, AA, 243, 118

\bibitem[Salo \& Laurikainen(1993)]{salo93} Salo, H. \& Laurikainen, E.
1993, ApJ, 410, 586.

\bibitem[Samurovi\' c \& Danziger(2005)]{Samurovic2005} Samurovi\' c,  S. \&
Danziger, I. J. 2005, MNRAS, 363, 769

\bibitem[Sansom et al.(2000)]{Sansom00} Sansom, A. E., Hibbard, J. E.
\& Schweizer, F. 2000, AJ, 120, 1946

\bibitem[Sarazin (1997)]{Sarazin97} Sarazin, C. L. 1997, The Nature of
  Elliptical Galaxies, ASP Conf.\ Ser.\ 116, M. Arnaboldi, G. S. Da Costa \&
  P. Saha Eds., San Francisco: ASP, 375

\bibitem[Sersic (1968)]{Ser68} Sersic, J. L. 1968, Atlas de Galaxias Australes,
Osservatorio Astronomico, Cordoba

\bibitem[Sulentic et al.(2006)]{Sulentic06} Sulentic, J. W. et al. 2006, AA,
  449, 937

\bibitem[Thomas et al.(2003)]{tho03} Thomas, D., Maraston, C. \&
Bender, R., 2003, Astrophys.
\& Space Science, 281, 371

\bibitem[Thomas et al.(2005)]{thom05} Thomas, D., Maraston, C.,
Bender, R. \& de Oliveira, C. M. 2005, ApJ, 621, 674

\bibitem[Tonry \& Davies(1981)]{TD81} Tonry, J. \& Davies, M. 1981, ApJ, 246, 666

\bibitem[Treu et al.(2005)]{Treu05} Treu, T. et al. 2005, ApJ, 622, L5

\bibitem[Trinchieri \& Rampazzo(2001)]{TR01} Trinchieri, G. \& Rampazzo, R.
2001, AA, 374, 454

\bibitem[Trinchieri et al.(2003)]{Tri03} Trinchieri, G., Sulentic, J.,
Breitschwerdt, D. \& Pietsch, W. 2003, AA, 401, 173

\bibitem[Tully (1988)]{Tully88} Tully, B.R. 1988, Nearby Galaxy
Catalogue, Cambridge University Press

\bibitem[Valdes (1998)]{vald98} Valdes, F.\ G.\ 1998, Astronomical Data
Analysis Software and Systems VII, ASP Conf.\ Ser.\ 145, R. Albrecht,
R. N. Hook, \& H. A. Bushouse Eds., San Francisco: ASP, 7

\bibitem[Verdes-Montenegro et al.(2005)]{lvm05} Verdes-Montenegro, L.,
Sulentic, J., Lisenfeld, U., Leon, S., Espada, D., Garcia, E., Sabater, J.
\& Verley, S. 2005, AA, 436, 443

\bibitem[Vikhlinin et al.(1999)]{Vik99} Vikhlinin et al. 1999, ApJ,
520, 1

\bibitem[Visvanathan \& Sandage(1977)]{Vis77} Visvanathan, N. \& Sandage, A.
1977, ApJ, 216, 214


\bibitem[Wilms et al. (2000)]{wilm} Wilms, J., Allen, A. \& McCray, R.
2000, ApJ, 542, 914

\bibitem[Zabludoff \& Mulchaey(1998)]{ZM98} Zabludoff, A. \& Mulchaey, J.
1998, ApJ, 496, 39

\bibitem[Zabludoff (1999)]{Z99}  Zabludoff, A. 1999, IAU Symp. 192,
eds. P. Whitelock and R. Cannon, 433

\bibitem[Zacharias et al.(2000)]{zach+00} Zacharias, N., et al.\ 2000,
\aj, 120, 2131
\end{thebibliography}
\end{document}